\documentclass[10pt,journal]{IEEEtran}
\usepackage{amsmath,amsfonts}
\usepackage{algorithmic}
\usepackage{algorithm}
\usepackage{array}
\usepackage[caption=false,font=normalsize,labelfont=sf,textfont=sf]{subfig}
\usepackage{textcomp}
\usepackage{stfloats}
\usepackage{url}
\usepackage{verbatim}
\usepackage{graphicx}
\usepackage{cite}
\usepackage{verbatimbox}
\usepackage{fancyvrb}
\usepackage{lipsum}

\newcommand{\Figure}[1]{Figure~\ref{#1}}
\newcommand{\Figures}[2]{Figures~\ref{#1} and \ref{#2}}

\newcommand{\Equation}[1]{Equation~\eqref{#1}}

\newcommand{\Table}[1]{Table~\ref{#1}}

\newcommand{\Section}[1]{Section~\ref{#1}}
\newcommand{\Sections}[2]{Sections~\ref{#1}~and~\ref{#2}}

\usepackage{pgfplots}
\pgfplotsset{compat=newest}
\usetikzlibrary{plotmarks}
\usetikzlibrary{arrows.meta}
\usepgfplotslibrary{patchplots}

\newcommand{\citep}[1]{\cite{#1}}
\newcommand{\citet}[1]{\cite{#1}}

\newcommand{\subparagraph}{} 
\usepackage{titlesec}

\newif\iforiginal

\originalfalse  

\begin{document}


\title{Towards Improved Objective Perceptual Audio Quality Assessment - Part 1: A
Novel Data-Driven Cognitive Model}

\author{Pablo M. Delgado and Jürgen Herre,~\IEEEmembership{Senior Member,~IEEE}
\thanks{Manuscript received March 18, 2024; revised July 27, 2024.}
\thanks{Pablo M. Delgado is with the Fraunhofer Institute for Integrated Circuits IIS,
91058 Erlangen, Germany (e-mail: pablo.delgado@iis.fraunhofer.de).}
\thanks{Jürgen Herre is with International Audio Laboratories
Erlangen, a joint institution of the Friedrich Alexander Universität Erlangen-Nürnberg,
and Fraunhofer IIS, 91058 Erlangen, Germany, and also with the Fraunhofer Institute for Integrated Circuits IIS, 91058 Erlangen, Germany (e-mail: juergen.herre@audiolabs-erlangen.de).}

\thanks{Digital Object Identifier 10.1109/TASLP.2021.XXXXXXX}
}

\markboth{IEEE/ACM TRANSACTIONS ON AUDIO, SPEECH, AND LANGUAGE PROCESSING, VOL. XX, 2023}{Shell \MakeLowercase{\textit{et al.}}: A Sample Article Using IEEEtran.cls for IEEE Journals}


\maketitle

\begin{abstract}
Efficient audio quality assessment is vital for streamlining audio codec development. Objective assessment tools have been developed over time to algorithmically predict quality ratings from subjective assessments, the gold standard for quality judgment. Many of these tools use perceptual auditory models to extract audio features that are mapped to a basic audio quality score prediction using machine learning algorithms and subjective scores as training data. However, existing tools struggle with generalization in quality prediction, especially when faced with unknown signal and distortion types. This is particularly evident in the presence of signals coded using non-waveform-preserving parametric techniques. Addressing these challenges, this two-part work proposes extensions to the Perceptual Evaluation of Audio Quality (PEAQ - ITU-R BS.1387-1) recommendation. Part 1 focuses on increasing generalization, while Part 2 targets accurate spatial audio quality measurement in audio coding.

To enhance prediction generalization, this paper (Part 1) introduces a novel machine learning approach that uses subjective data to model cognitive aspects of audio quality perception. The proposed method models the perceived severity of audible distortions by adaptively weighting different distortion metrics. The weights are determined using an interaction cost function that captures relationships between distortion salience and cognitive effects. Compared to other machine learning methods and established tools, the proposed architecture achieves higher prediction accuracy on large databases of previously unseen subjective quality scores. The perceptually-motivated model offers a more manageable alternative to general-purpose machine learning algorithms, allowing potential extensions and improvements to multi-dimensional quality measurement without complete retraining.

\end{abstract}

\begin{IEEEkeywords}
Objective audio quality assessment, machine learning, perception, psychoacoustics, Perceptual Evaluation of Audio Quality (PEAQ), ITU-R BS.1387-1, ViSQOL.
\end{IEEEkeywords}

\section{Introduction}
\label{sec:intro}


The development of audio coding technologies has been a thriving field of research for the three last decades, particularly since the standardization of the MPEG-1 Audio Layer III (MP3) format \cite{brandenburg1994iso}. With each generation, audio codecs enable the transmission of larger volumes of high-quality communication and media content to a growing audience \cite{OverviewMediaCompression2021}. The pursuit of greater efficiency and improved audio quality continues to be a top priority in the ongoing quest for advancements in codec development. Given the rapid pace of technological advancement, there is a pressing need for codec development to have a quick time to market. In this context, objective quality assessment -- also known as quality \textit{measurement} -- assumes a crucial role: by predicting subjective ratings in overall audio quality assessment tasks, it saves time and resources associated with subjective quality assessment, considered to be the ultimate quality reference. To accomplish this, the majority of the established objective quality assessment methods make use of models of auditory perception \cite{RixTaslp, PEAQ,POLQAcite,ViSQOLAudio, PEMOQ}. Perceptual objective quality assessment has served various purposes, ranging from selection and tuning of perceptual audio codecs for specific applications to optimization and monitoring of transmission networks considering customer quality-of-experience (QoE) \cite{RixTaslp}. More recently, objective tools have demonstrated their usefulness in ablation studies to facilitate a better understanding of the relationship between design choices and resulting audio quality in the development of neural audio codecs \cite{defossez2022high}. Additionally, perceptual quality metrics have recently been employed as loss functions in training deep noise suppression networks as an alternative to Mean Squared Error (MSE) losses (see e.g., \cite{FingscheidtPESQ}). This approach has led to enhanced quality compared to other baseline methods.

\begin{figure*}[htb]
    \begin{center}
      \centering
      \resizebox{\width}{!}{
          \includegraphics[width=2\columnwidth]{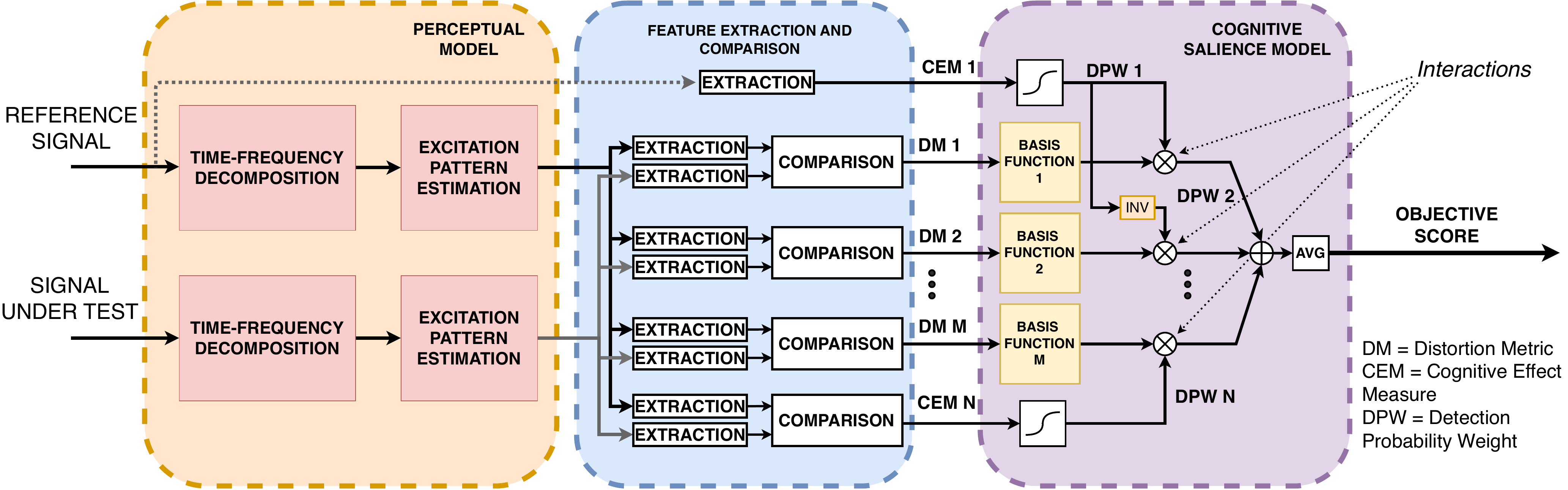}
      }

    \end{center}
    \caption{Block diagram of the proposed objective quality assessment system. The perceptual model corresponds to that used in the PEAQ method \citet{PEAQ}. The feature extraction stage contemplates the existing PEAQ features (advanced version), as distortion metrics in the time-pitch-loudness domain through the excitation patterns. The proposed extension uses additional features for the effect size measurement of different cognitive phenomena. The novel cognitive salience model structure constitutes an explicit interaction model of cognitive effects and distortion metrics.}
    \label{fig:block_diagram}
\end{figure*}

A highly sought-after attribute of an objective quality assessment system is the ability to predict quality degradation reliably across diverse quality levels and application scenarios, emphasizing the need for robust generalization \cite{TorcoliApplication}. Multi-dimensional perceptual quality measurement systems seek to address this issue by using several separate perceptually-motivated distortion metrics (DMs) that model multiple aspects of quality degradation (i.e., dullness, roughness, noisiness), each dominant in different quality ranges and applications \cite{PEAQ, POLQAcite, PESQ}. These systems use machine learning (ML) approaches to map the different DMs to a single quality score using subjective responses of overall quality assessment tasks as training data. This mapping stage can be seen as an application-specific cognitive model of auditory perception \cite{ThiedePEAQ, RixPhD}. However, the use of general-purpose ML algorithms can produce overfitted models if the training data is scarce, leading to reduced prediction accuracy outside of the training set. Deep-learning based quality metrics have also been proposed, primarily targeting reference-free speech quality \cite{SoniDeepSpeechQuality}. However, more recently, metrics for audio quality have emerged \cite{NetFlix, manocha2020differentiable, jiang2023generative}. The challenge with deep-learning approaches lies in the substantial data required for training, and these methods are currently in the validation phase. 

In \cite{delgado2022data}, we showed a novel way of incorporating subjective data into an improved cognitive model that features a perceptually-motivated ML architecture. We argued that objective quality prediction generalization can be improved by extending the quality mapping stage of a multi-dimensional quality measurement system to incorporate cognitive effect metrics (CEMs). These CEMs depend on the nature of the input signal, and they have been shown to predict the importance of certain DMs (and the irrelevance of others) in estimating the quality of different cases. Therefore, the CEM outputs can be used as adaptive weights to the DMs. This method is akin to modeling the influence that cognitive effects have on the salience \cite{bregman1994auditory} of audible distortions. In initial tests, this approach has demonstrated improved generalization performance with the typical orders of magnitude of training data utilized in training objective metrics.

Our work builds on \cite{delgado2022data} in several ways. In \Section{sec:method}, we provide a detailed explanation of the proposed method and its assumptions. The PEAQ method is extended to include CEMs and its original shallow neural network mapping stage is replaced with a new cognitive model. The novel cognitive model training incorporates subjective data in two different stages: first, mapping of the available DMs to a subjective scale, and second, modeling of the interactions between quality metrics and cognitive effects using an interaction cost function.

The method is validated on a wide range of audio coding technology quality scores. The description of the model calibration and validation procedure is described in \Section{sec:Experiment} and our findings are presented in \Section{sec:results}. The results show that our ML model performs better in generalization of quality prediction than other general ML models when using the same training data, and also generalizes better than other state of the art quality measurement methods for the analyzed data.

In \Section{sec:discussion}, we discuss how certain aspects of the presented method agree with results in previous studies on audio quality perception. Particularly, we discuss the improved prediction accuracy shown in some aspects, such as quality prediction of parametrically coded audio signals, where other methods have had limited success.

From a ML point of view, the proposed method addresses the overfitting problem by reducing training noise and reducing the amount of training parameters. Namely, 1) it reduces training noise by estimating quality mapping functions separately for each DM with a listening test database containing isolated audio artifacts and 2) it reduces the number of training parameters by limiting predictor interactions to meaningful cognitive effect and DMs.

Additionally, the presented perceptually-motivated model architecture provides a more tractable model than those resulting from other general-purpose ML algorithms by separating the data-driven modeling of interactions between cognitive effects and distortion metrics from the quality mapping. This tractable structure potentially allows for additional extensions and improvements based on further perceptual rules without the need to completely retrain the model each time a new type of distortion is caused by newer audio processes. We hope that our approach to incorporate subjective quality data into multi-dimensional quality assessment systems will inspire others in the audio quality metric development community, leading to more comprehensive and effective objective quality assessment methods.

\section{Method}
\label{sec:method}
The proposed audio quality measurement system, illustrated in \Figure{fig:block_diagram}, aims to predict a subjective grade in the form of a mean quality score of overall perceived audio quality over many subjects, by comparing undisturbed references (REF) and potentially degraded signals under test (SUT) using models of human hearing and cognition \cite{RixTaslp}. This work will focus on the prediction of subjective scores collected with the Multi-Stimulus Test with Hidden Reference and Anchor (MUSHRA) methodology \cite{MUSHRA} and, to a minor extent, scores collected with the procedure described in the ITU-R BS.1116 recommendation \cite{BS1116}. However, the presented method can potentially generalize to similar subjective quality assessment procedures.

The perceptual model analyzes input signals by applying psychoacoustically-motivated transformations to capture auditory processes relevant to the quality assessment task. It calculates DMs that measure various aspects of quality degradation in the psychophysical domain \cite{letowski1989sound}. These metrics are usually informed with data from isolated psychoacoustic experiments. Additionally, the model computes the so-called CEMs to represent different effect sizes of cognitive auditory phenomena. While the DMs mainly consider peripheral hearing effects, the CEMs capture central auditory processes that influence the perception of quality degradation. These cognitive effects impact the salience of distortions and contribute to the overall perception of quality degradation.

The CEMs and DMs are combined using a Cognitive Salience Model (CSM) in order to produce the objective quality score \cite{delgado2022data}. The CSM replaces the multivariate regression models found in other objective quality assessment systems for the quality mapping stage\cite{PEAQ,flener2017assessment, DelgadoPEAQ, barbedo2005a, RixPhD, SpeechEvaluationMultipleForeground}. The outputs of each DM are transformed from the psychophysical domain to a perceived quality degradation scale through the basis functions and then combined with a weighted summing scheme. The relative weights are regulated by the effect sizes measured by the CEMs and represent an explicit model of cognitive effect/distortion salience interaction, emulating the influence of cognitive phenomena on the salience of particular degradations. The time-dependent weighted sum can then be averaged over the whole duration of the signals in order to produce a single quality score for the signal comparison. Additionally, the CEM outputs are mapped into the quality domain using sigmoid functions, to account for possible effect detection thresholds and non-linearities. The CSM parameters are informed by subjective listening test data. Therefore, listener preference is also expected to be modeled in this stage.

\subsection{Perceptual Model}
For the perceptual model we use our MATLAB \cite{MATLAB} implementation \citep{DelgadoPEAQ} of the model used by the PEAQ method \citep{PEAQ}, in its \textit{advanced} version \footnote{Although this MATLAB implementation is not publicly available, similar results could be achieved with publicly available implementations such as the one presented in \cite{holters2015gstpeaq}.}. The perceptual model performs a comparison of REF and SUT (e.g., potentially degraded by a perceptual audio codec) in the transformed psychophysical domain \cite{PESQ, POLQAcite, beerends1992a, Zwickerbook}. After a pre-processing stage \footnote{Pre-processing consists of time-alignment and silence and DC removal \cite{PEAQ}, omitted in \Figure{fig:block_diagram} for clarity.}, the front-end stage performs the transformation from the physical (time-frequency-intensity) to the psychophysical (time-pitch-loudness or amplitude modulation) domain, which involves a time-frequency decomposition to account for the signal transformations taking place in the basilar membrane \cite{ThiedePHD}. Among the most important aspects modeled by this stage are phenomena related to simultaneous and non-simultaneous masking \cite{Zwickerbook}, which are fundamental in perceptual audio coding technology. The transformed representations are usually termed the \textit{internal representations} \citep{beerends1992a} or \textit{excitation patterns}. The excitation patterns of REF and SUT are compared using different distance metrics from which the DMs are derived in the feature extraction stage. The perceptually-motivated DMs have been reported to show a stronger correlation to subjective quality degradation judgement scores than other non-perceptual signal quality measures calculated in the physical domain that do not take into account perceptual factors (e.g., signal-to-noise ratio or mean squared error \cite{RixTaslp}). Further features describing different effect sizes of cognitive phenomena can also be extracted from the excitation patterns.

\subsubsection{Distortion Metrics (DMs)}
\label{sec:dm_metrics}
\begin{table}[t]
  \centering
  \caption{Used distortion metrics and their related PEAQ MOVs (Tables 4 and 17 in \citet{PEAQ}).}
\resizebox{\columnwidth}{!}{%
\begin{tabular}{|l|c|c|}
    \hline
    \textbf{DM} & \textbf{Associated MOV} &\textbf{Distortion Degradation Type} \\ \hline
    LinDist & AvgLinDist\_A & Linear distortions  \\
    ModDiff & RmsModDiff\_A & Modulation disturbances \\
    NoiseLoudness & RmsNoiseLoud\_A & Added noise in SUT \\
    MissingComponents & RmsMissingComponents\_A & Missing components in SUT \\
    EHS  & EHS\_B & Harmonic structure of the error \\
    SegNMR  & Segmental NMR\_B & Noise-to-mask ratio \\
    \hline
\end{tabular}
}
\label{tab:distmetrics}
\end{table}

The DMs from the perceptual model are calculated in the same manner as PEAQ's Model Output Values (MOVs) of the advanced version, excluding the time-averaging steps. The overall time averaging step is carried out after the weighted sum in the CSM. The used DMs and their associated MOVs are shown in Table \ref{tab:distmetrics}.

\subsubsection{Cognitive Effect Metrics (CEMs)}
\label{sec:cog_metrics}
\begin{table}[t]
\caption{Used cognitive effect metrics.}
  \centering
      \resizebox{\columnwidth}{!}{%
      \begin{tabular}{|l|c|}
        \hline
        \textbf{CEM} & \textbf{Description} \\ \hline
        EPN & Perceptual streaming measure from \citep{barbedo2005a}, Eq. (30)  \\
        PDEV & Informational masking from \citep{barbedo2005a}, Eq. (31) \\
        probSpeech & Probability of speech-like signal from \citep{USACSpeechClas} \\
        \hline
      \end{tabular}
    }
	\label{tab:cognimetrics}
\end{table}

In addition to DMs, CEMs (Table \ref{tab:cognimetrics}) can be derived from the perceptual model. The CEMs describe phenomena in quality perception beyond masking effects that affect the perceived severity of a given distortion.

In the field of Auditory Stream Analysis \citep{bregman1994auditory}, two important phenomena for perceiving audio quality degradation are Perceptual Streaming (PS) and Informational Masking (IM) \citep{bregman1994auditory}. PS refers to the extraction of meaningful elements from complex auditory scenes, where distortions perceived as separate events are easier to detect than those integrated with the signal \citep{what_to_listen_for_2}. Separate percepts are considered more annoying and impact quality degradation more significantly \citep{beerends1996the}. Conversely, IM occurs when a disturbance above the masking threshold becomes undetectable or less prominent due to increased masker complexity (in this context, the coded signal is considered the masker). Disturbances with the same loudness may be less noticeable or bothersome in signals with rapid and large variations compared to those present in signals with smaller and slower variations. PS and IM counteract each other, we implemented CEMs for these phenomena as described in \citep{barbedo2005a}, calculated on the excitation patterns of REF and SUT.

Speech and music quality perception involve slightly different auditory nerve processing mechanisms \citep{SpeechVsMusicQuality}. Therefore, a speech-music classifier from \citep{USACSpeechClas} is used as an example CEM on the REF signal to determine the likelihood of a signal being speech or music. Optimized for low-complexity and low-latency in the context of audio coding, this classifier analyzes signals after silence removal but before psychophysical representation transformations, as represented in the upper CEM branch of \Figure{fig:block_diagram}. Other classifiers based on the input excitation patterns may also be used. The classifier extracts features based on Perceptual Linear Prediction Cepstral Coefficients (PLPCC), voicing strength, and pitch to inform a Gaussian Mixture Model-based classifier. The decisions are computed every 16 ms of audio signal and later synchronized with the system's overall temporal granularity of 100 ms.

\subsection{Cognitive Salience Model (CSM) Architecture}
\label{sec:CSM}

As explained, the CSM depicted in \Figure{fig:block_diagram} operates under the assumption that cognitive effects regulate the salience of competing distortion phenomena. This assumption is modeled in the CSM through the interactions between DMs and CEMs within an adaptive weighted sum scheme. The transformation process from cognitive and distortion measures into an objective score is described below.

\subsubsection{Quality Basis Functions}
\label{sec:BF}

To perform the weighted sum, individual DMs from the perceptual model are mapped to a target quality scale using basis functions (BFs). The BFs account for compression effects and scale biases between DMs and subjective scores \citep{RixPhD}. The BFs are estimated through a regression procedure with a training database of subjective scores on degraded signals with a wide range of perceived quality. The database used to obtain the BFs is presented in \Section{sec:calibration_db}.

The BFs are estimated using Multivariate Adaptive Regression Splines (MARS) \citep{Jekabsons_areslab}. The default hyperparameters of the MATLAB implementation in \citep{Jekabsons_areslab} were used, except for the maximum number of included MARS basis functions (maxFuncs), which was set to 3 to ensure monotonically decreasing mappings \citep{RixPhD, ThiedePHD}. The transformed DMs in the subjective quality domain (denoted as DMs with subscript Q) represent the output of BFs on the MUSHRA scale.

\subsubsection{Detection Probability Weights}
\label{sec:DPW}

The influence of a certain cognitive effect on the salience of a given distortion can be subject to certain nonlinear aspects. For example, a cognitive effect may only influence the perception of a given distortion above a certain threshold, or this influence can also saturate after a certain effect size. We model this phenomenon using psychometric functions \citep{PsychometricCitation}. These functions effectively capture the detection probability of the cognitive phenomena mentioned in \Section{sec:cog_metrics} that are interacting with the different types of distortion perception described by the different DMs of \Section{sec:dm_metrics}. To model the detection probability of the cognitive effects, the psychometric functions are applied to the CEMs outputs and termed the \textit{Detection Probability Weights} (DPWs). The DPWs parameters for each meaningful CEM/DM interaction are estimated separately. In cases where an increasing cognitive effect size diminishes the salience of a distortion artifact type (e.g., increasing IM decreases the perceived severity of a certain distortion characterized by one DM, e.g., noise loudness), an inverse operation in the detection probability domain can be represented as $1-DPW$.

\subsection{Interaction Model of Cognitive Effects and Distortion Salience}
\label{sec:interaction_model}

\begin{figure}[htb]
  \begin{center}
    \centering
    \resizebox{\columnwidth}{!}{%
        \includegraphics[width=0.6\textwidth]{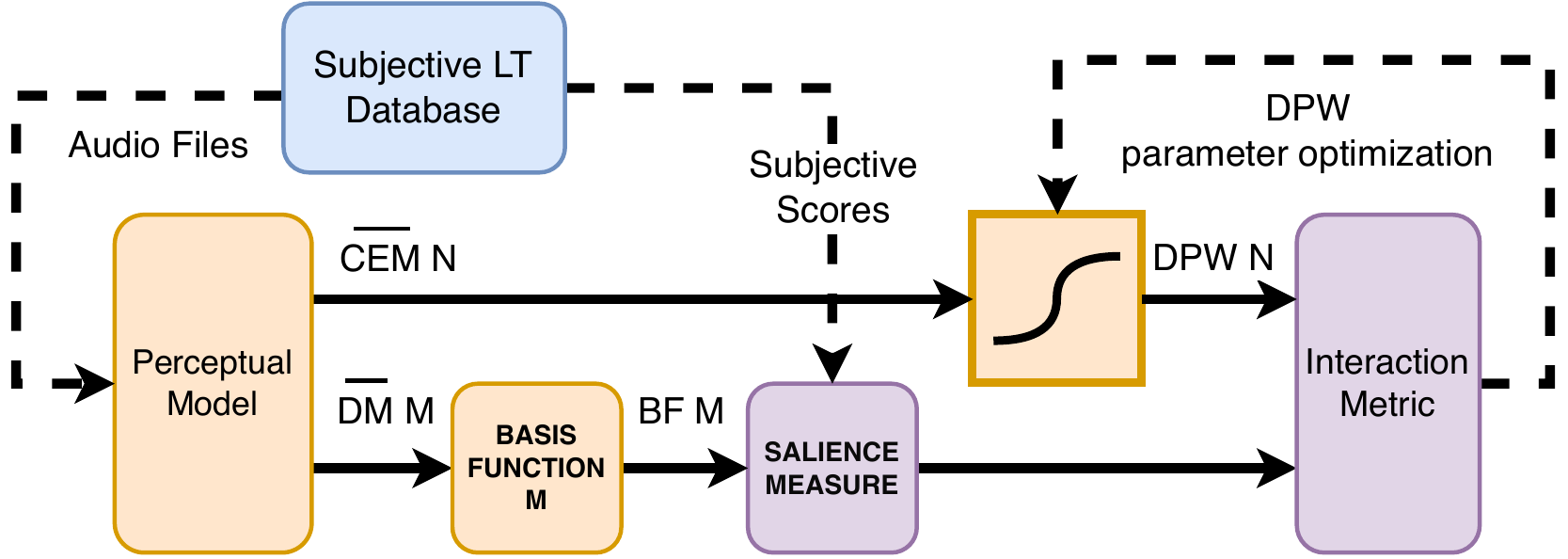}
    }
\end{center}
\caption{Block diagram for the interaction selection and optimization procedure.}
\label{fig:optimization}
\end{figure}

The meaningful interactions between CEM and the salience of the difference distortions are determined using an iterative optimization process (\Figure{fig:optimization}).

The procedure searches for the parameters of the psychometric functions that maximize the cost function values, using the outputs of the perceptual model on an extensive database of perceptually coded signals and their subjective quality scores. The associated cost function is an interaction metric based on the correlation between the CEM outputs, after transformation through the DPWs, and the estimated salience of the distortions associated to the DM outputs. As target data, the optimization process uses listening test scores and audio material from the database described in \Section{sec:TrainingAndValidData}. The CEM/DM salience interactions with the strongest interaction metric values will be kept as connections in the CSM model. For the calculation of the interaction metric, an estimate of the distortion salience is needed.

\subsubsection{Salience Measure}
\label{sec:salience_measure}
It is assumed that a distortion type will be salient if its associated DM can predict quality degradation. In order to provide an estimate of said DM's prediction power, a measure of correlation between DM and mean subjective scores can be used. The salience $\mathcal{S}$ for a DM $m,\mbox{ }m=1\ldots M$ in a signal $j, \mbox{ } j=1\ldots J$ is defined as the Pearson correlation coefficient between the respective DM basis function $BF_m$ and the mean subjective score $y$ for all available treatments \footnote{ i.e., a signal is treated/processed by a particular audio codec.} $i=1\ldots I$:
\begin{equation}
  \label{eq:salience}
  \resizebox{0.9\columnwidth}{!}{
  $\mathcal{S}_m(j)
   = \frac{\sum_{i=1}^{I} ( y_{ij}-\overline{y}_{j} )( BF_{mij}-\overline{BF}_{mj} )}
  {\sqrt{\sum_{i=1}^{I} (y_{ij}-\overline{y}_{j})^2} \sqrt{\sum_{i=1}^{I} (BF_{mij}-\overline{BF}_{mj})^2}}
  $}.
\end{equation}
The metric is calculated for averaged values over time for the duration of the signals. A strong correlation suggests that the DM effectively reflects the perceived quality degradation (i.e., the associated distortion is more salient), while a weak correlation indicates that the DM might not adequately describe the impact on quality.

\subsubsection{Interaction Metric}
\label{sec:interaction_metric}
The interaction metric describes the covariance of the salience metric for a DM against the CEM (transformed by the DPW) across all signals in the database. This metric quantifies DM/CEM interaction effects: 

\begin{equation}
  \label{eq:costFunction}
  \resizebox{0.9\columnwidth}{!}{$
  \mathcal{C}_m
    = \bigg|\frac{\sum_{j=1}^{J} ( S_m(j)-\overline{S}_{m} )( DPW_{m}(j)-\overline{DPW}_{m} )}
  {\sqrt{\sum_{j=1}^{J} (S_{m}(j)-\overline{S}_{m})^2} \sqrt{\sum_{j=1}^{J} (DPW_{m}(j)-\overline{DPW}_{m})^2}}\bigg|
  $}
\end{equation}

The stronger the correlation, the better the CEM will predict a given DM salience over the signals in the database. Note that either strong positive or negative covariance (or correlation) can denote prediction power in this case. Additionally, two DPW sigmoid parameters (curve steepness and crossover midpoint) are chosen in an exhaustive search procedure in order to maximize the interaction metrics. The extreme values of the sigmoid function are kept between 0 and 1 to reflect a detection probability of the associated cognitive effect.

\subsubsection{Interaction Selection}
\label{sec:parameter_selection}
The CEM/DM salience pairs with the strongest interaction metrics are selected using input variable importance analysis. For this, linear regression with step-wise selection is used to maximize the adjusted $R^2$ metric \citep{hastie2009elements}, which penalizes increased model complexity. Interactions and higher order terms are not allowed as they are already explicitly modeled by $\mathcal{C}_m$. Predictors are centered and scaled using z-score scaling. The resulting regression coefficients represent the importance of associated quality terms. Only important quality terms are included in the final cognitive model. Each candidate DPW multiplied by its target BF is an independent quality predictor.

\section{Calibration and Validation Procedure}
\label{sec:Experiment}

\begin{table*}[ht]
    \caption{Experimental Aspect Summary of the Subjective Audio Quality Score Databases Used for Calibration and Validation.}
\centering
  \resizebox{0.85\width}{!}{%
  \begin{tabular}{|p{20mm}|p{20mm}|p{20mm}|p{22mm}|p{22mm}|p{17mm}|p{17mm}|p{17mm}|p{17mm}|}
    \hline
    \textbf{Label} $\rightarrow \rightarrow$ & \textbf{MPEG-H}\textsuperscript{1} & \textbf{ITU DB4}\textsuperscript{2} & \textbf{ELD VT (A)}\textsuperscript{3} & \textbf{ELD VT (T)}\textsuperscript{3} &  \textbf{USAC VT 3}\textsuperscript{4} & \textbf{USAC VT 2}\textsuperscript{4}& \textbf{USAC VT 1}\textsuperscript{4$\dagger$} & \textbf{Isolated Artifacts}\textsuperscript{5$\dagger$} \\ \hline
    \textbf{Content Type}  & \scriptsize{Music, dialog, \newline soundscapes, sound effects.}        & \scriptsize{Music, dialog, \newline soundscapes, sound effects.} & \scriptsize{Music}, \newline \scriptsize{speech (clean, background noise, reverberant)}, \scriptsize{mixed.} & \scriptsize{Music, speech, mixed.} & \scriptsize{Music, speech, mixed.} & \scriptsize{Music, speech, mixed.} & \scriptsize{Music, speech, mixed.} & \scriptsize{Music, speech, mixed.} \\ \hline
    \textbf{Channel Format}  & 22.1         & 5.1 & Stereo & Stereo & Stereo & Stereo & Mono & Stereo \\ \hline
    \textbf{Methodology}    & MUSHRA & BS.1116 & MUSHRA & MUSHRA & MUSHRA & MUSHRA & MUSHRA & MUSHRA \\ \hline
    \textbf{Mean Quality}\textsuperscript{6} & 80.5 & -0.5 & 81.7 & 67.9 & 73.7 & 57.1 & 58.3  & 66.3 \\ \hline
    \textbf{Codecs} & \scriptsize{MPEG-H, Channel and Object-based} & \scriptsize{MPS, Dolby AC3, HE-AAC (v2 core), AAC-LD, AAC-ELD} & \scriptsize{AAC-ELD,HE-AAC} \newline AAC-LD, G.722 & \scriptsize{AAC-ELD,HE-AAC} \newline AAC-LD, G.722 & \scriptsize{USAC, AMR-WB+, HE-AACv2} & \scriptsize{USAC, AMR-WB+, HE-AACv2} & \scriptsize{USAC, AMR-WB+, HE-AACv2} & \scriptsize{Modified HE-AAC and other audio processes} \\ \hline
    \textbf{Bitrate (kbps)} & 256-1200 & 128-320 & 32-64 & 24-32 & 32-96 & 16-24 & 8-24 & -- \\ \hline
    \textbf{N (Mean Scores)} & 144 & 280 & 231 & 156 & 192 & 168 & 216 & 200 \\ \hline
    \multicolumn{8}{l}{\textsuperscript{1}\footnotesize{ISO/IEC JTC1/SC29/WG11 N13633\citet{MPEGHdatabase}}, \textsuperscript{2}\footnotesize{ITU-R 6C/415-E \citet{ITUDB45}}, \textsuperscript{3}\footnotesize{ISO/IEC JTC1/SC29/WG11 N10032 \citet{ELDdatabase}}, \textsuperscript{4}\footnotesize{ISO/IEC JTC1/SC29/WG11 N12232 \citet{USACdatabase}},} \\
    \multicolumn{8}{l}{\textsuperscript{5}\footnotesize{Dick et at. \cite{dick2017generation}} \textsuperscript{6}\footnotesize{All values expressed in MUSHRA scores\cite{MUSHRA}, except ITU DB4, expressed in Subjective Difference Grade (SDG)\cite{BS1116}.} \textsuperscript{$\dagger$}\footnotesize{Calibration database}.} \\
    \end{tabular}
    }
    \label{tab:table_subj}
\end{table*}

\begin{table}[ht]
    \caption{Summary of the Blind Source Separation Subjective Audio Quality Database.}
\centering
  \resizebox{0.85\width}{!}{%
  \begin{tabular}{|p{20mm}|p{20mm}|}
    \hline
    \textbf{Label} $\rightarrow \rightarrow$ & \textbf{SEBASS-DB (SASSEC)}\textsuperscript{7} \\ \hline
    \textbf{Content Type} & \scriptsize{Music (incl.percussion), speech} \\ \hline
    \textbf{Channel Format}  & Stereo  \\ \hline
    \textbf{Methodology} & MUSHRA  \\ \hline
    \textbf{Mean Quality} & 35.1 \\ \hline
    \textbf{Algorithms} & \scriptsize{11 different BSS algorithms\textsuperscript{8}.} \\ \hline
    \textbf{N (Mean Scores)} & 154 \\ \hline
    \multicolumn{2}{l}{\textsuperscript{7}\footnotesize{Kastner et al.\cite{SEBASS}}, \textsuperscript{8}\footnotesize{Vincent et al.}\cite{vincent2007first}} \\
  \end{tabular}
}
    \label{tab:table_subj_SASSEC}
\end{table}

Calibration and validation are crucial to establish the precision and dependability of quality prediction systems. Calibration adjusts the parameters of the CSM architecture to accurately mimic subjective responses from a reference (i.e., calibration) database. This includes calculating the BFs, identifying significant interactions between cognitive effects and distortion salience, and optimizing DPW parameters to yield accurate objective scores mirroring subjective quality assessments. Conversely, validation compares the calibrated system's objective scores (and its variants) with subjective scores of new, unseen data to verify the system's reliability.

\subsection{Performance Metrics}

During the validation process of a quality measurement system, system performance is usually rated in terms of correlation $R$ between predictor (objective scores) and target (subjective scores). Particularly, we used Pearson's correlation coefficient to evaluate system linearity performance \citep{EvalObjective}. For the validation data, we additionally tested other metrics such as prediction error and the number of outliers. However, these will not be reported as the performance ranking remained consistent across all measures; in other words, the systems with the strongest $R$ values also exhibited the lowest prediction errors and number of outliers.

\subsection{Subjective Audio Quality Data}
\label{sec:TrainingAndValidData}

For the most part of the calibration and validation, we used listening test databases produced during audio codec and quality measurement system standardization efforts \cite{MPEGHdatabase, ELDdatabase,ITUDB45,USACdatabase}. The main experimental aspects of the databases are summarized in \Table{tab:table_subj}. All listening tests were carried out according to the MUSHRA procedure, with the exception of the the ITU-R WP/6C Training Database for the Revision of PEAQ -- ITU DB4 \cite{ITUDB45}, which was carried out using the quality assessment procedure in the ITU-R BS.1116 recommendation \cite{BS1116}.

We utilized an additional validation dataset to illustrate the model's usability beyond the realm of audio coding. This dataset is a subset of SEBASS-DB \cite{SEBASS}, a comprehensive public database of listening test results for evaluating Blind Source Separation (BSS) quality. Specifically, it includes results from listening tests conducted with signals from the stereo audio source separation evaluation campaign (SASSEC) \cite{vincent2007first}. Further details are shown in \Table{tab:table_subj_SASSEC}.

These databases are desirable for system validation because they are large (in terms of number of subjective scores collected and listening panel size) and diverse (in terms of present distortion artefacts, signal types, coding technologies and quality ranges). It is assumed that the performance metrics calculated over large and diverse datasets will generalize to further unseen data. Therefore, the associated system will predict subjective responses with a similar reliability after deployment.

\subsection{Calibration Database}
\label{sec:calibration_db}

For estimating the BF, we used the subjective listening test database presented in \citep{dick2017generation} which primarily contains signals degraded with isolated audio coding artifacts (i.e, the \textit{isolated artifacts} database). A subjective quality database collected on isolated artifacts is desirable for the estimation of the basis functions, as the interactions between different degradation dimensions are minimized.

However, the isolated artifacts database is not ideal for estimating CEM/DM interactions in the CSM because the audio treatments only contain one type of distortion at a time. For determining meaningful CEM/DM interactions, we use the Unified Speech and Audio Codec Verification Test 1 database -- USAC VT1 \cite{USACdatabase} (\Table{tab:table_subj}). The USAC VT1 features signals processed with different coding technologies at different bitrates, and with audio treatments that contain multiple competing distortions for different signal types. The separation into balanced sets of speech, music and mixed signals is particularly useful for the training of the DPW associated with the $probSpeech$ CEM.

\subsection{Validation Databases}
All of the databases listed in \Table{tab:table_subj} and \Table{tab:table_subj_SASSEC}, except for the isolated artifacts database, are used for system validation.  With the exception of USAC VT1, none of the validation databases were used for calibrating any of the parameters of the proposed quality measurement system.

\subsection{Mapping Stage Validation}
\label{sec:proposed}

Besides evaluating the performance of a quality measurement system that uses the CSM, the validation procedure also considers different alternative methods for combining the CEMs of \Table{tab:cognimetrics} and DMs of \Table{tab:distmetrics} into a single objective score (i.e., mapping stages). These alternatives reflect previously adopted approaches for incorporating cognitive effects into quality metrics. 

$\mathbf{BASELINE}\mbox{ }\mathbf{DM}\mbox{ }\mathbf{(ANN)}$: An objective system that exclusively calculates the DMs listed in \Table{tab:distmetrics}. The DMs are mapped to mean subjective scores but no CEMs are used. The mapping stage consists of an Artificial Neural Network (ANN) with the same settings as reported for the advanced version of PEAQ \citep{RevisionBS1387}. This system provides the baseline performance for the case in which no CEMs influence the DMs.

$\mathbf{DM + CEM\mbox{ }(ANN)}$, $\mathbf{DM + CEM\mbox{ }(KSOM)}$ and $\mathbf{DM + CEM\mbox{ }(MARS)}$: The labels describe objective systems using different ML algorithms that map DMs and CEMs to an objective score. An ANN with the same settings as in $BASELINE$ is used for the $DM + CEM (ANN)$ case. Additionally, emulating the approach used in \citet{barbedo2005a}, the same features are mapped to an objective quality score using a Kohonen self-organizing map (KSOM) classifier. The built-in KSOM implementation in MATLAB is used with the default parameters with a self organizing map of dimensions $3 \times 3000$ to reach a sufficient level of granularity in the classification. The classes were then monotonically mapped to the MUSHRA scale with full resolution. This configuration has shown to produce the best prediction performance of the model for the training and validation data. Additionally, a MARS learning model is also included in the comparison, as MARS models have been reported to show a good performance in objective quality assessment systems \citep{Flessner}. The MARS model that showed the best results was allowed a maximum number of basis functions to be 16 (two times the number of input features), and rest of the initialization parameters were taken as recommended in \citep{Jekabsons_areslab}. 

\textbf{PEAQ-CSM} and \textbf{PEAQ-CSM+}: Two system variants incorporating the methods described in \Section{sec:method} are evaluated. PEAQ-CSM implements the CSM proposed in \Section{sec:CSM}, combining CEMs and the BFs of \Section{sec:BF} as a weighted sum, without using the DPW. The system PEAQ-CSM+ consists of the PEAQ-CSM system that additionally uses the DPW. The performance difference between the two systems should indicate if there is any gain in modeling detection probability and listener preference for cognitive effects with the DPW.

\subsection{Pre-processing of Audio Files}

All the audio files were delay-compensated based on the estimated time-delay from the maximum value of the cross-correlation function between REF and SUT. As for playback level adjustment, the PEAQ-based methods follow the procedure described in \citet{PEAQ}, Section 2.2.3. The rest of the methods assume a playback level of 100 dB SPL at full scale (i.e., at values of $\pm 1$ in MATLAB signals). The signals are scaled to a presentation level of around 65 dB SPL and resampled to 48 kHz, if necessary. Additionally, noise and silence before and after the signals are removed according to Section 5.2.4.4 of the BS.1387 recommendation \citep{PEAQ}. This process also truncates the evaluated signal duration so that they enter the quality measurement systems with the same length. The multi-channel signals in the MPEG Evaluation Listening Test for 3D Audio -- MPEG-H \cite{MPEGHdatabase} and ITU DB4 database have been mixed down to a binaural signal using the Binaural Room Impulse Response of the Fraunhofer IIS ``Mozart" listening room described in \citet{silzle2009vision}.

\subsection{Comparison to State of the Art}
\label{sec:stateofartsystems}

As a part of the system validation procedure, a series of state-of-the-art quality measurement methods will also be evaluated in terms of quality prediction performance, namely:

\begin{itemize}
    \item \textbf{CombAQ} corresponds to a MATLAB implementation \citep{CombAQGit} of the quality metric proposed in \citet{flener2019}. The metric combines a binaural audio quality metric \citep{flener2017assessment} and a timbre quality metric \citep{biberger2018objective}.
    \item \textbf{GPSMq} corresponds to the timbre quality metric output \citep{biberger2018objective} of the MATLAB implementation of CombAQ.
    \item \textbf{PEMO-Q} is an implementation in MATLAB \citep{emiya2010peass} of the PEMO-Q metric presented in \citet{PEMOQ}. The comparison considers the $PSMt_{min}$ metric outputs with the built-in modulation filter bank analysis (FB) and without modulation filter bank analysis (LP). The inclusion of the modulation filterbank in the quality estimation is expected to improve accuracy.
    \item \textbf{ViSQOL Audio} is a MATLAB implementation (rev. 238, \citet{Visqol_soft}) of the objective quality assessment system discussed in \citet{ViSQOLAudio}. The metric ViSQOL NSIM is calculated in full-band settings for the music files and in wide-band configuration for speech files. This configuration provided the best performance in preliminary tests.
    \item \textbf{PEAQ DI} is a MATLAB implementation of the objective measure specified in \citet{PEAQ}, used in \textit{advanced} operation mode. The MATLAB implementation was validated against the output of a commercially available implementation, as reported in \citet{DelgadoPEAQ}.
  \end{itemize}

\section{Results}
\label{sec:results}
This section presents the results of the calibration and validation stages of the proposed quality measurement system. The results refer to the data-driven CSM parameter selection method presented in \Sections{sec:CSM}{sec:interaction_model}.

\subsection{Basis Function Estimation}

\begin{figure}[htb]
\begin{center}
  \centering
  \resizebox{\columnwidth}{!}{
      \input{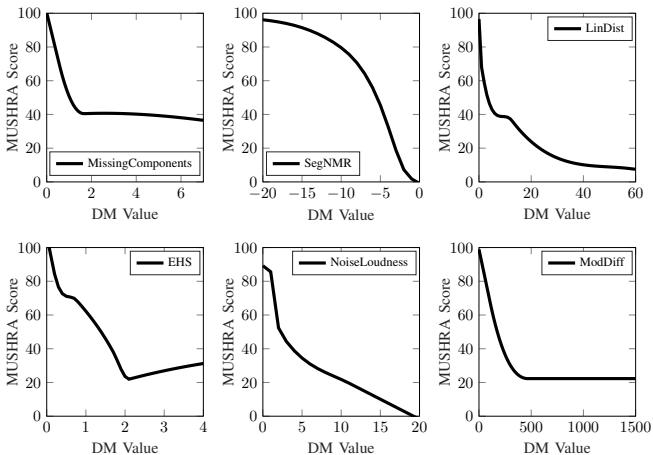}
  }
\end{center}
\caption{Basis Functions mapping individual DM outputs to quality scores estimated with the procedure described in \Section{sec:BF}.}
\label{fig:Basis_Functions}
\end{figure}

\Figure{fig:Basis_Functions} shows estimates of the basis functions from \Section{sec:BF}, corresponding to each DM listed in \Table{tab:distmetrics}. Generally, these estimated basis functions show a predominantly monotonically decreasing trend, where larger DM values indicate lower quality. One exception is the BF associated with $EHS$, which does not strictly decrease for DM values of 2 and above. However, this deviation is not likely to significantly impact the overall estimation performance. It occurs in the lower quality range of the scale, where a greater variance of subjective quality ratings than in the higher quality range is usually observed.

\subsection{Cognitive Effect and Distortion Salience Interaction Selection}
\label{sec:interaction_selection}
\begin{table*}[t]
\caption{Candidate interactions for the CSM and interaction metric $\mathcal{C}$ values (before optimization/after optimization) for the calibration database USAC Verification Test 1.}
\centering
  \resizebox{\width}{!}{%
    \begin{tabular}{|l|c|c|c|c|}
        \hline
      \textbf{Weight} & \textbf{CEM} &\textbf{Target BF} & $\mathbf{\mathcal{C}}$& \textbf{Equation} \\
      \hline
      DPW1    & probSpeech & LinDist\_Q            & -0.77/-0.92    & DPW1 = 1-probSpeech\_th\_lin \\
      DPW2    & probSpeech & NoiseLoudness\_Q      &  0.67/0.80     & DPW2 = probSpeech\_th\_nl \\
      DPW3    & probSpeech & MissingComponents\_Q  & -0.60/-0.89    & DPW3 = 1-probSpeech\_th\_mc \\
      DPW4    & probSpeech & EHS\_Q                &  0.70/0.71     & DPW4 = 1-probSpeech\_th\_ehs \\
      DPW5    & EPN        & LinDist\_Q            & -0.40/-0.70    & DPW5 = 1-EPN\_th\_lin \\
      DPW6    & EPN        & EHS\_Q                &  0.57/0.60     & DPW6 = EPN\_th\_ehs \\
      DPW7    & PDEV       & EHS\_Q                & -0.67/-0.67    & DPW7 = 1-PDEV\_th\_ehs \\
      DPW8\_a & EPN        & SegNMR\_Q             &  0.1 / 0.25    & DPW8 = (EPN\_th\_sgm)(1-PDEV\_th\_sgm) \\
      DPW8\_b & PDEV       & SegNMR\_Q             & -0.18 / -0.21  & Incorporated into DPW8 \\
      DPW9\_a & EPN        & NoiseLoudness\_Q      &  0.32 / 0.34   & DPW9 = (EPN\_th\_nl)(1-PDEV\_th\_nl) \\
      DPW9\_b & PDEV       & NoiseLoudness\_Q      & -0.52 / -0.52  & Incorporated into DPW9 \\
      \hline
    \end{tabular}
  }

  \label{tab:candidate_DPW}
\end{table*}

Results of the procedure described in \Section{sec:interaction_model} are presented here. \Table{tab:candidate_DPW} shows the CEM/DM interaction candidates for the CSM. The $\_th$ suffix indicates the ``thresholded” CEM with the optimized logistic function, considering the target DM salience. For reference, the column $\mathcal{C}$ shows the values of the interaction metrics (i.e., the cost function) before and after DPW parameter optimization.

\begin{figure}[htb]
  \begin{center}
    \centering
      \resizebox{0.85\columnwidth}{!}{
%
%
\definecolor{mycolor1}{rgb}{0.00000,0.44700,0.74100}%
\definecolor{mycolor2}{rgb}{0.92900,0.69400,0.12500}%
\begin{tikzpicture}

\begin{axis}[%
width=10.5cm,
height=9cm,
scale only axis,
xmin=0.5,
xmax=1,
xlabel style={font=\Large},
xlabel={$\text{DM Salience (LinDist)}$},
ymin=0,
ymax=1.2,
ylabel style={font=\Large},
ylabel={Mean Effect Size},
axis background/.style={fill=white},
axis x line*=bottom,
axis y line*=left,
xmajorgrids,
ymajorgrids,
legend style={at={(0.57,0.30)}, anchor=north east, legend cell align=left, align=left, draw=white!15!black, font=\large},
tick label style={font=\fontsize{10}{12}\large}, 
axis line style={thick, line width=1pt}, 
grid style={thick, line width=0.8pt}, 
]
\addplot[only marks, mark=*, mark size=4pt, mark options={}, fill=mycolor1,draw=mycolor1] table[row sep=crcr]{%
x	y\\
0.921689302794165	0.519395131092052\\
0.690160528153531	0.958839894090783\\
0.729118520088729	0.979663764717109\\
0.886560187174597	0.814959876840239\\
0.994806672119575	0.488100864019267\\
0.925168209407104	0.416183762332928\\
0.963829766644456	0.651583451017625\\
0.859210810397561	0.535197609970899\\
0.988942751866792	0.617504375565135\\
0.920970097431821	0.626242420612211\\
0.881739276965217	0.747848879016157\\
0.580492226661912	1\\
0.805825143543113	0.925263261034612\\
0.88405334788852	0.344208429137637\\
0.824079981189414	0.829953743813112\\
0.966951726037525	0.635674114173775\\
1	0.588729858126584\\
0.954168647734673	0.520380099534952\\
0.953912559558891	0.543266163908604\\
0.942357624626653	0.374970945867713\\
0.87367200006372	0.679891498709342\\
0.83288991848072	0.863845670062008\\
0.662591544624924	0.985800618470931\\
0.926104265700772	0.372320376492107\\
};
\addlegendentry{probSpeech S: -0.77}

\addplot [color=black, thick]
  table[row sep=crcr]{%
0	1.9561678816668\\
0.0434782608695652	1.89204056932667\\
0.0869565217391304	1.82791325698653\\
0.130434782608696	1.76378594464639\\
0.173913043478261	1.69965863230625\\
0.217391304347826	1.63553131996611\\
0.260869565217391	1.57140400762597\\
0.304347826086957	1.50727669528583\\
0.347826086956522	1.4431493829457\\
0.391304347826087	1.37902207060556\\
0.434782608695652	1.31489475826542\\
0.478260869565217	1.25076744592528\\
0.521739130434783	1.18664013358514\\
0.565217391304348	1.122512821245\\
0.608695652173913	1.05838550890486\\
0.652173913043478	0.994258196564726\\
0.695652173913043	0.930130884224587\\
0.739130434782609	0.866003571884449\\
0.782608695652174	0.80187625954431\\
0.826086956521739	0.737748947204171\\
0.869565217391304	0.673621634864033\\
0.91304347826087	0.609494322523894\\
0.956521739130435	0.545367010183756\\
1	0.481239697843617\\
};
\addlegendentry{propForSpeech fit}

\addplot[only marks, mark=diamond*, mark options={}, mark size=4pt, draw=black, fill=gray] table[row sep=crcr]{%
x	y\\
0.921689302794165	0.00375051999439402\\
0.690160528153531	0.817700672454315\\
0.729118520088729	0.913936606593578\\
0.886560187174597	0.219773054924863\\
0.994806672119575	0.00229448721181233\\
0.925168209407104	0.000613508418611306\\
0.963829766644456	0.0254966711340117\\
0.859210810397561	0.00476605811702463\\
0.988942751866792	0.0157783771041201\\
0.920970097431821	0.0178570106276827\\
0.881739276965217	0.094531537780277\\
0.580492226661912	1\\
0.805825143543113	0.654039695778106\\
0.88405334788852	0\\
0.824079981189414	0.26170208117847\\
0.966951726037525	0.0203977369088841\\
1	0.0104576822082636\\
0.954168647734673	0.00380748586915404\\
0.953912559558891	0.00537756556111127\\
0.942357624626653	0.000189730935207951\\
0.87367200006372	0.0377719553722926\\
0.83288991848072	0.378214198051791\\
0.662591544624924	0.940851212929987\\
0.926104265700772	0.000169970816173378\\
};
\addlegendentry{$\text{probSpeech (opt.) }\text{S: -0.92}$}

\addplot [color=gray, /tikz/dashed]
  table[row sep=crcr]{%
0	2.8040325989981\\
0.0434782608695652	2.67575538258334\\
0.0869565217391304	2.54747816616858\\
0.130434782608696	2.41920094975382\\
0.173913043478261	2.29092373333907\\
0.217391304347826	2.16264651692431\\
0.260869565217391	2.03436930050955\\
0.304347826086957	1.90609208409479\\
0.347826086956522	1.77781486768003\\
0.391304347826087	1.64953765126528\\
0.434782608695652	1.52126043485052\\
0.478260869565217	1.39298321843576\\
0.521739130434783	1.264706002021\\
0.565217391304348	1.13642878560625\\
0.608695652173913	1.00815156919149\\
0.652173913043478	0.879874352776729\\
0.695652173913043	0.751597136361971\\
0.739130434782609	0.623319919947213\\
0.782608695652174	0.495042703532455\\
0.826086956521739	0.366765487117697\\
0.869565217391304	0.238488270702939\\
0.91304347826087	0.110211054288181\\
0.956521739130435	-0.0180661621265772\\
1	-0.146343378541335\\
};
\addlegendentry{$\text{propForSpeech (opt.) fit}$}

\end{axis}
\end{tikzpicture}%
      }
  \end{center}
  \caption{Mean Effect Size for a CEM (speech probability) and DM salience (linear distortions) for the signals in the USAC Verification Test Database 1. The DPW transformation increases DM/CEM covariance given by $\mathcal{C}$ as calculated by \Equation{eq:costFunction}.}
  \label{fig:plot_CEM_vs_MOV}
  \end{figure}
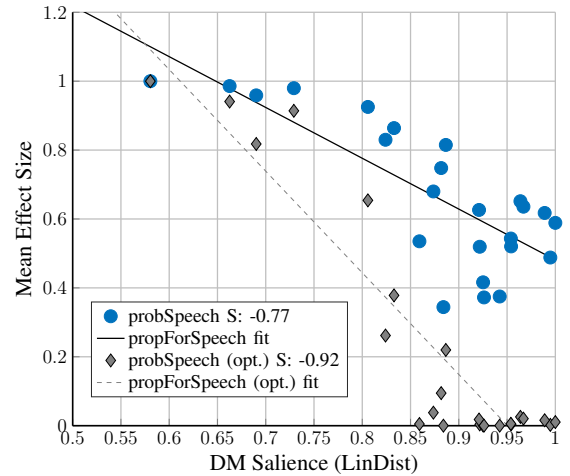

Candidates $DPW1$ to $DPW7$ were chosen due to strong resulting values of the data-driven interaction metric. For example, a greater probability of the signal being speech-like ($probSpeech$) is expected to predict harmonic distortion ($EHS$) salience to a certain degree ($\mathcal{C} = 0.71$). A negative sign indicates lower DM salience for a larger effect size. For instance, the higher the probability that a signal is speech-like, the less salient linear distortions ($LinDist$) are expected to be ($\mathcal{C} = -0.92$, see \Figure{fig:plot_CEM_vs_MOV}).

\begin{table}[t]
  \caption{Selected final model parameters for the CSM of \Figure{fig:block_diagram} based on step-wise linear regression results on the optimization database. The initial candidate quality predictors were listed in \Table{tab:candidate_DPW}.}
\centering
  \resizebox{\columnwidth}{!}{%
    \begin{tabular}{|l|c|c|c|}
      \hline
      \textbf{Quality Term} & \textbf{Expression} & \textbf{Associated CEM} & \textbf{Reg. Coef} \\
      \hline
      $Q_0$ & Intercept                &  none         &    58.3  \\
      $Q_1$ & $DPW1*LinDist\_Q$        &  probSpeech   &    2.69  \\
      $Q_2$ & $DPW2*NoiseLoudness\_Q$  &  probSpeech   &    2.61  \\
      $Q_3$ & $DPW4*EHS\_Q$            &  probSpeech   &    1.97  \\
      $Q_4$ & $DPW5*LinDist\_Q$        &  EPN          &    1.54  \\
      $Q_5$ & $DPW6*EHS\_Q$            &  EPN          &    1.64  \\
      $Q_6$ & $DPW9*NoiseLoudness\_Q$  &  EPN/PDEV     &   -1.89  \\
      $Q_7$ & $RmsModDiff\_Q$          &  none         &    8.49  \\
      \hline
    \end{tabular}
  }

  \label{tab:importanceOptim}
\end{table}

In addition, \Table{tab:candidate_DPW} considers other candidate interactions based on previous literature linking cognitive effects with distortion severity perception. The candidate parameters $DPW8$ and $DPW9$ incorporate combined effects of PS and IM into the two distortion metrics from Table I associated with added noise, $SegNMR$ and $NoiseLoudness$, in a similar fashion to the method reported in \citet{beerends1996the}.

The results of the step-wise regression described in \Section{sec:parameter_selection} are summarized in \Table{tab:importanceOptim}. Since no CEMs have been shown to strongly interact with $RmsModDiff\_Q$, no DPW will be associated with this quality term. The terms associated with the candidate interaction of $DPW8$ showed extremely weak $\mathcal{C}$ values and have been therefore dropped by the regression procedure. All estimated regression coefficients have associated p-values, $p < 0.05$ at a confidence level of $95\%$. The final goodness-of-fit measures for the regression are $R=0.91$ for correlation of target mean subjective scores against model outputs and a root mean squared error $RMSE=6.18$. Both measures indicate that the model accurately describes the target data and the importance analysis can be taken into consideration.

\begin{figure}[htb]
  \begin{center}
  \resizebox{\columnwidth}{!}{%
    \resizebox{0.9\columnwidth}{!}{
    \raggedleft
      \input{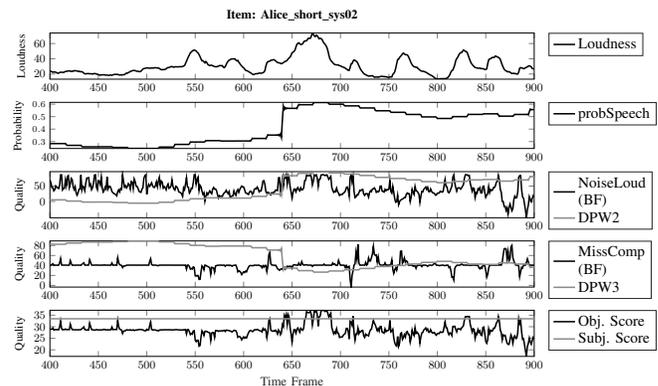}
    }
  }
  \end{center}
\caption{An illustrative CSM model output for an audio item of the USAC Verification Test 1 database. The signal is composed of an orchestral music excerpt followed by female speech, around frame 640. The CSM model steers the objective score output in time towards the salient DM basis function output according to the probability of the signal being speech-like. Top plot: reference signal loudness. Second-to-top: probability of speech-like signal. Middle plot: noise loudness BF outputs and corresponding DPW. Second-to-bottom plot: missing component loudness BF output and corresponding DPW. Bottom Plot: CSM output over time and the mean MUSHRA subjective score for the analyzed item. Temporal granularity is $t_s = 0.1 s$ per frame.}
\label{fig:CH_timbre__combined_metric_time}
\end{figure}

The CSM of \Figure{fig:block_diagram} with the final parameters will produce a quality metric $\mathcal{QM}$ of the form:
\begin{equation}
  \mathcal{QM}(n) = \sum_{k=0}^7 Q_k(n)
   \label{eq:sum_Q}
\end{equation}
where $Q_k$ are the quality terms as described in \Table{tab:importanceOptim} and $n$ is the time index in samples. A mean quality metric $\overline{\mathcal{QM}}$ over all time samples (see e.g., \Figure{fig:CH_timbre__combined_metric_time}) is chosen as an estimate of the mean subjective score of the potentially degraded signal under test.

\subsection{Mapping Stage Validation and Comparison to State of The Art}
\label{sec:validation}

\begin{figure}[ht]
\resizebox{0.45\textwidth}{!}{%

\begin{tikzpicture}
\begin{axis}[
    axis on top,
    xticklabels={USAC VT1\textsuperscript{$\dagger$}, USAC VT3, ELD VT (A), ELD VT (T), MEAN},
    yticklabels={BASELINE DM (ANN), DM + CEM (ANN), DM + CEM (KSOM),DM + CEM (MARS), PEAQ-CSM, PEAQ-CSM+},
    xlabel={Listening Test DB},
    ylabel={Mapping Stage},
    xlabel style={yshift=-1em},
    ylabel style={yshift=1em},
    xtick=data,
    ytick=data,
    colormap={mycolormap}{
        color=(red),
        color=(red),
        color=(red!30!white),
        color=(green),
    },
    colorbar,
    point meta min=0, 
    point meta max=1, 
    colorbar style={
        xlabel=R,
        xlabel style={yshift=-1em}
    },
    xticklabel style={rotate=90}, 
    nodes near coords,
    nodes near coords style={
        font=\small,
        font=\bfseries,
        anchor=center,
    },
    point meta=explicit,
    point meta rel=per plot,
]
\addplot[
    matrix plot*,
    mesh/cols=5,
    point meta=explicit,
    mesh/rows=6
] coordinates {

    (0,1) [0.90]
    (1,1) [0.76]
    (2,1) [0.71]
    (3,1) [0.42]
    (4,1) [0.69]

    (0,2) [0.91]
    (1,2) [0.74]
    (2,2) [0.73]
    (3,2) [0.28]
    (4,2) [0.67]

    (0,3) [0.80]
    (1,3) [0.72]
    (2,3) [0.71]
    (3,3) [0.56]
    (4,3) [0.71]

    (0,4) [0.91]
    (1,4) [0.70]
    (2,4) [0.64]
    (3,4) [0.35]
    (4,4) [0.65]

    (0,5) [0.88]
    (1,5) [0.76]
    (2,5) [0.79]
    (3,5) [0.76]
    (4,5) [0.79]

    (0,6) [0.90]
    (1,6) [0.81]
    (2,6) [0.85]
    (3,6) [0.83]
    (4,6) [0.86]
};
\end{axis}
\end{tikzpicture}
}
\caption{Validation of the CSM as a mapping stage in terms of objective/subjective score correlation $R$ performance (see. \citep{EvalObjective}), and performance comparison to the alternative mapping stages listed in Section \ref{sec:proposed}. Confidence intervals: $CI_{95\%_R} \leq \pm 0.01$ for all estimates. $\dagger$: \footnotesize{Used as training data in ANN and MARS mapping stages, but only for CEM-DM salience interaction and DPW parameter selection in PEAQ-CSM/CSM+}.}
\label{tab:heatmap_R}
\end{figure}
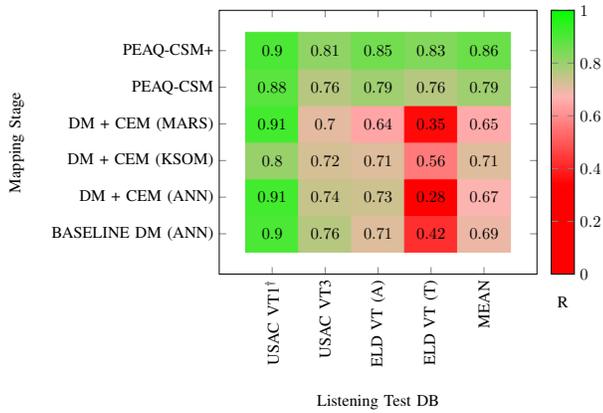

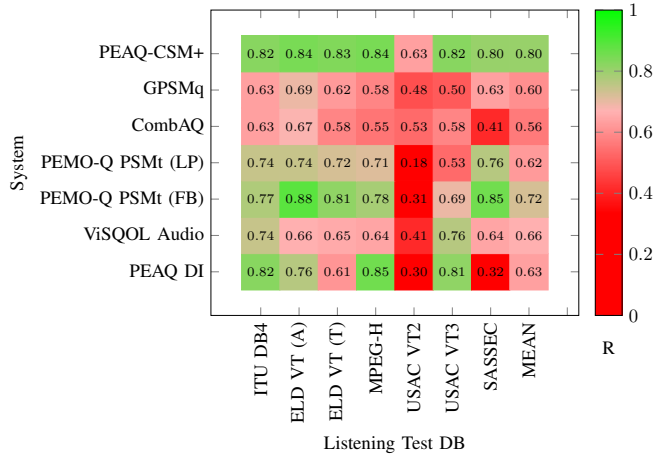
\begin{figure}[ht]
\resizebox{0.49\textwidth}{!}{%
\begin{tikzpicture}
\begin{axis}[
  colorbar,
  colorbar style={
    xlabel=R,
    xlabel style={yshift=-1em}
  },
  colormap={mycolormap}{
    color=(red),
    color=(red),
    color=(red!30!white),
    color=(green),
  },
  point meta min=0, 
  point meta max=1, 
  xlabel={Listening Test DB},
  ylabel={System},
  xtick={1,...,9},
  xticklabels={
    ITU DB4,
    ELD VT (A),
    ELD VT (T),
    MPEG-H,
    USAC VT2,
    USAC VT3,
    SASSEC,
    MEAN
  },
  ytick={1,...,7},
  yticklabels={
    PEAQ DI,
    ViSQOL Audio,
    PEMO-Q PSMt (FB),
    PEMO-Q PSMt (LP),
    CombAQ,
    GPSMq,
    PEAQ-CSM+,
  },
  yticklabel style={align=right},
  xticklabel style={rotate=90}, 
  nodes near coords style={
    font=\small,
    font=\bfseries,
    anchor=center,
},
]

\addplot[
  matrix plot*,
  mesh/cols=8,
  mesh/rows=7,
  point meta=explicit,
  nodes near coords={\pgfmathprintnumber[fixed zerofill,precision=2]{\pgfplotspointmeta}},
  nodes near coords style={font=\scriptsize},
] coordinates {
(1, 1)  [0.82] 
(2, 1)  [0.76] 
(3, 1)  [0.61] 
(4, 1)  [0.85] 
(5, 1)  [0.3]  
(6, 1)  [0.81] 
(7, 1)  [0.32] 
(8, 1)  [0.63] 

(1, 2)  [0.74]
(2, 2)  [0.66]
(3, 2)  [0.65]
(4, 2)  [0.64]
(5, 2)  [0.41]
(6, 2)  [0.76]
(7, 2)  [0.64] 
(8, 2)  [0.66]

(1, 3)  [0.77]
(2, 3)  [0.88]
(3, 3)  [0.81]
(4, 3)  [0.78]
(5, 3)  [0.31]
(6, 3)  [0.69]
(7, 3)  [0.85] 
(8, 3)  [0.72]

(1, 4)  [0.74]
(2, 4)  [0.74]
(3, 4)  [0.72]
(4, 4)  [0.71]
(5, 4)  [0.18]
(6, 4)  [0.53]
(7, 4)  [0.76] 
(8, 4)  [0.62]

(1, 5)  [0.63]
(2, 5)  [0.67]
(3, 5)  [0.58]
(4, 5)  [0.55]
(5, 5)  [0.53]
(6, 5)  [0.58]
(7, 5)  [0.41] 
(8, 5)  [0.56]

(1, 6)  [0.63]
(2, 6)  [0.69]
(3, 6)  [0.62]
(4, 6)  [0.58]
(5, 6)  [0.48]
(6, 6)  [0.5]
(7, 6)  [0.63] 
(8, 6)  [0.60]

(1, 7)  [0.82]
(2, 7)  [0.84]
(3, 7)  [0.83]
(4, 7)  [0.84]
(5, 7)  [0.63]
(6, 7)  [0.82]
(7, 7)  [0.80] 
(8, 7)  [0.80]

};

\end{axis}
\end{tikzpicture}
}
\caption{System validation in terms of objective/subjective score correlation ($R$) performance \citep{EvalObjective} for the systems listed in \Section{sec:stateofartsystems}. $CI_{95\%} \leq \pm 0.01$ for all estimates.}
\label{tab:heatmap_overall_R}
\end{figure}

The results for the validation procedure described in \Section{sec:proposed} and the comparison to other state of the art quality measurement systems of \Section{sec:stateofartsystems} are shown in  \Figures{tab:heatmap_R}{tab:heatmap_overall_R}, respectively.

In terms of mapping stage validation, the CSM-based mapping stages outperformed alternative methods for all validating databases, except for the USAC VT1 case. In this case, system predictions using ANN and MARS only show a strong correlation against subjective scores because USAC VT1 has been used as training data. In contrast, both CSM variants were the only ones that show a consistent prediction power across unseen databases.

On the other hand, PEAQ-CSM+, the CSM model version using DPWs, demonstrated superior results for all databases compared to the version without DPWs, PEAQ-CSM. This indicates the effectiveness of functional thresholding in modeling detection probabilities, nonlinearities, and listener dependence on perceived degradation severity.

In evaluating the $DM+CEM$ mapping stages relative to the baseline and those systems that do not use CEMs, our findings underscore that the employment of CEM is warranted solely in conjunction with the proposed CSM. Incorporating additional CEM as input variables into the assessed general-purpose ML algorithms, without explicit modeling of cognitive effect and distortion salience interactions, yielded unsatisfactory results.

\Figure{tab:heatmap_overall_R} also indicates that other established objective audio quality metrics failed to predict quality degradations with satisfactory linearity performance\footnote{A correlation coefficient of $R > 0.8$ is normally considered satisfactory in the literature \cite{ThiedePHD, RixPhD}} as consistently across databases as the proposed method. A more detailed discussion about these aspects is given in the next Section.

\begin{figure*}
  {\centering \textbf{Listening Test DB: ELD VT (A)} \par\medskip}
  \subfloat[PEAQ-CSM versus PEAQ-CSM+.\label{fig:OPTsfig}]{
  \resizebox{0.99\columnwidth}{!}{
  \input{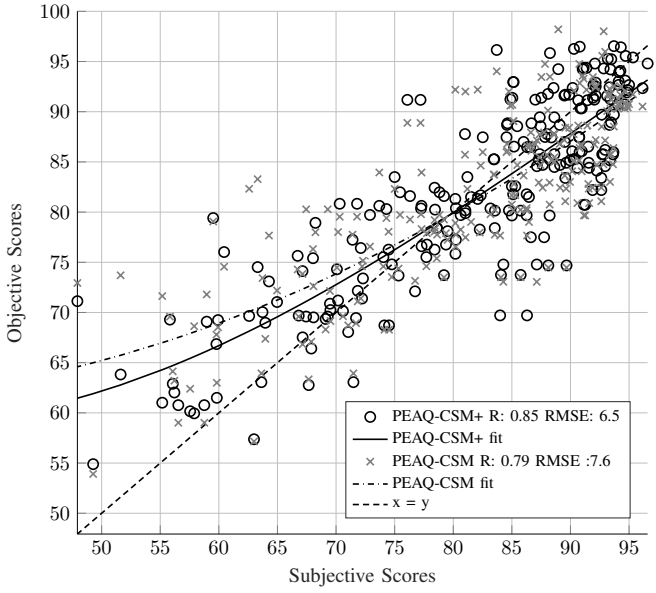}}
  }
  \hfill
  \subfloat[PEAQ-CSM+ versus ViSQOL(NSIM).\label{fig:ViSQOLsfig}]{
    \resizebox{0.99\columnwidth}{!}{
    \input{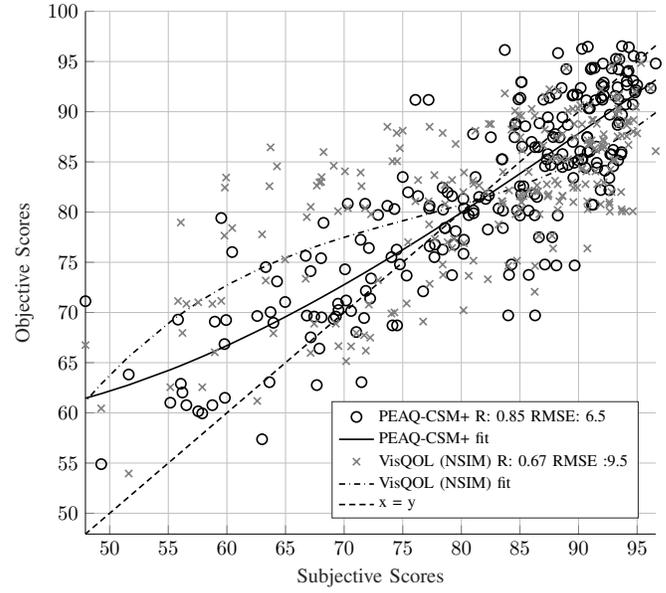}}
    }

  \subfloat[PEAQ-CSM+ versus $\mathbf{DM+CEM\mbox{ }(ANN)}$.\label{fig:ANNsfig}]{
  \resizebox{0.99\columnwidth}{!}{
  \input{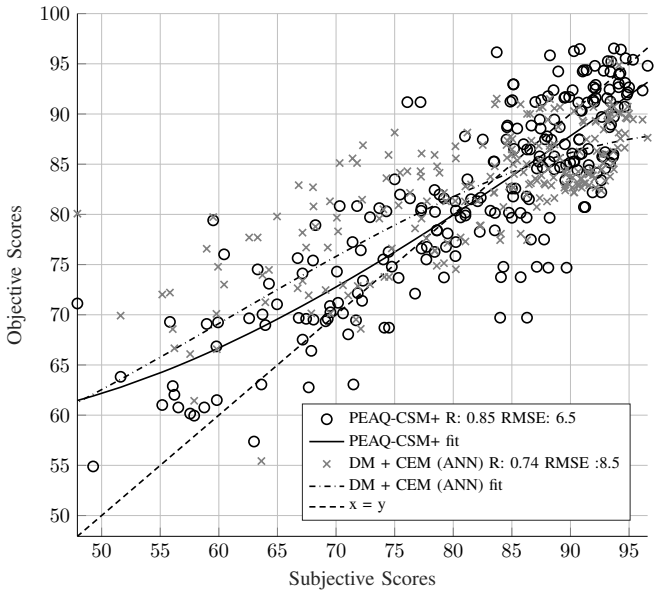}}
  }
  %
  \subfloat[PEAQ-CSM+ versus PEAQ's DI.\label{fig:PEAQsfig}]{
  \resizebox{0.99\columnwidth}{!}{
  \input{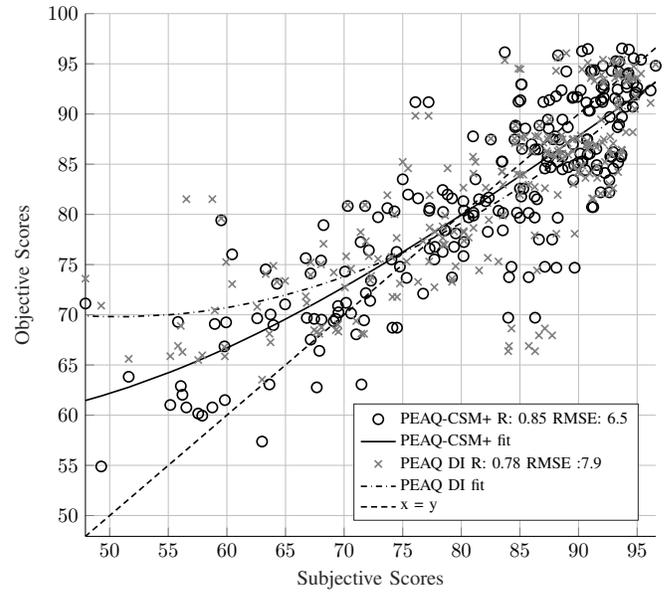}}
  }
  \hfill
  %

\caption{Objective versus subjective quality score scatter plots for PEAQ-CSM+ and four other systems described in \Section{sec:proposed} for the unseen Listening Test Database described in \citet{ELDdatabase}. The solid and dashed lines represent third order mapping functions applied to the system outputs in order to compensate scale biases and reduce the RMSE without changing rank order, as recommended in \citet{EvalObjective}. Performance metrics in terms of subjective versus objective score correlation (R) and RMSE for each considered system are shown in the legends.}
\label{fig:final_scatter}
\end{figure*}

\subsection{Inference Computational Complexity Estimation}
\label{sec:inference_complexity}
We also include a complexity analysis of the architectures described in \Section{sec:proposed}, in terms of number of parameters for each model and comparative mean inference time. The inference time measurements were carried out in MATLAB using a MacBook Pro 16" with a 2.6 GHz 6-Core Intel Core i7 Processor and 32 GB of RAM. Mean time measurements were taken out of 10000 training and inference cycles with data splitting and randomization on the USAC VT1 database, using the Monte Carlo procedure described in \cite{DelgadoPEAQ}. \Table{tab:table_complexity} presents the estimated inference complexity in terms of execution time and the number of parameters for the architectures detailed in \Section{sec:proposed}. 

\begin{table}[t]
  \caption{Comparative Inference Time and Model Complexity for the Architectures Presented in \Section{sec:proposed}.}
\centering
  \resizebox{\width}{!}{%
  \begin{tabular}{|p{40mm}|p{20mm}|p{10mm}|}
    \hline
    \textbf{Model} & \textbf{Mean Inference Time} & \textbf{No. of Param.} \\ \hline
    \textbf{$\mathbf{BASELINE}\mbox{ }\mathbf{DM}\mbox{ }\mathbf{(ANN)}$} & 3.60 ms &  36\\ \hline
    \textbf{$\mathbf{DM + CEM\mbox{ }(ANN)}$}  & 3.70 ms &  51 \\ \hline
    \textbf{$\mathbf{DM + CEM\mbox{ }(KSOM)}$}  & 15.1 ms &  48000\\ \hline
    \textbf{$\mathbf{DM + CEM\mbox{ }(MARS)}$} & 0.47 ms & 49 \\ \hline
    \textbf{PEAQ-CSM} & 0.41 ms & 37 \\ \hline
    \textbf{PEAQ-CSM+} & 0.51 ms & 59 \\ \hline
  \end{tabular}
}
  \label{tab:table_complexity}
\end{table}

\section{Discussion}
\label{sec:discussion}
The following Section discusses the reported results and their relationship with the different assumptions and statements that have been made during the development of the proposed quality measurement system.

\subsection{CSM Model Parameters: Agreement with Psychoacoustic Results}
\label{sec:csm_model_parameters}

\Sections{sec:interaction_model}{sec:interaction_selection} described how the CSM model parameter selection procedure employed two criteria to identify potential DM/CEM interactions. First, the DM salience had to significantly correlate with a related cognitive effect size. Second, the candidate interactions drew upon existing psychoacoustic literature. The final parameters were determined through a step-wise linear regression on subjective listening test data. The consistency of the selected interactions in \Table{tab:importanceOptim} with prior auditory perception research remains to be confirmed. In this respect, some observations on the final selected interactions of \Table{tab:importanceOptim} can be made:

$\mathbf{Q_1}$, $\mathbf{Q_2}$ and $\mathbf{Q_3}$: these terms mark the interactions of linear distortions, added noise loudness and harmonic-structured errors with the speech probability CEM. In $Q_1$, linear distortions are predicted to be more salient for non speech-like signals. Band limitation can be a significant driver of quality degradation in speech signals with multiple distortions \citep{SpeechEvaluationMultipleForeground}. However, listeners are expected to be more critical of band limitations in wider-bandwidth music signals than in speech signals when presented in the same test, as is the case in the USAC VT 1 database. As for $Q_2$, noise loudness-based metrics have been known to be a good predictor of quality on speech signals \citep{PESQ, POLQAcite, DelgadoPEAQ}. Therefore, the selection of these quality terms is supported by previous studies.

Regarding $Q_3$, errors with harmonic structures can elicit perceived roughness due to harmonic mismatch \citep{dick2017generation}, and these are not expected to be salient for speech signals exclusively. Although the $EHS$ feature has been reported to be a useful predictor for speech quality \cite{DelgadoPEAQ}, harmonic error signals are supposed to be salient in the case of music signals as well (e.g., bass clarinet, harpsichord, \citet{PEAQ}). A post-hoc analysis on music signals affected by ``harmonic mismatches" in the isolated artifacts database showed comparatively weak salience for $EHS$ ($\overline{S}_{EHS} = 0.85$, $\overline{S} = 0.92$ on all features). As these items were generated through bandwidth extension coding techniques, the results suggest that the $EHS$ DM is not reliable enough to measure quality degradation on music signals produced by these methods.

$\mathbf{Q_4}$ and $\mathbf{Q_5}$: Both quality terms, $Q_4$ and $Q_5$, tie DMs to the perceptual streaming impact of additional disturbances. $Q_4$ suggests that as perceptual streaming intensifies, the salience of linear distortions reduces. Linear distortions effectively predict quality degradation in the absence of other nonlinear artifacts. If additional disturbances start forming a separate percept, which expert listeners often critique (e.g. ``birdies" in \citet{what_to_listen_for_2, ArtifactsinPAC}), they can become dominant and therefore have a stronger effect in quality degradation than linear distortions. Meanwhile, $Q_5$ associates the heightened severity of harmonic errors with perceptual streaming, due to various spectral-pattern component groupings and local harmonicity effects potentially creating separate percepts \citet{HarmonicStructureGrouping}.

$\mathbf{Q_6}$: In the calibration process, the step-wise regression inversely polarized the quality term $Q_6$ related to disturbance loudness, compensating for other terms in \Equation{eq:sum_Q} that exaggerated signal distortion when subjective scores did not indicate such high degradation in  perceived quality. A post-hoc analysis indicated that this was mainly true for speech signals rated lower on the quality scale, near the $3.5$ kHz MUSHRA anchor. This suggests that $Q_6$ might be reacting to border effects in the subjective data and countering the positive polarity of $Q_2$, which is also related to disturbance loudness. The artifacts found in the signals are known as ``speech reverberation" or ``ghost voice" \citep{what_to_listen_for_2, herre_1996_tns}, and they may have caused extreme degradation predictions due to noise loudness in $Q_2$ when, in fact, they are not perceived as disturbing by the listeners. Further investigations are needed to better understand these dynamics.

\subsection{Usefulness of Modeling Nonlinearities in Cognitive Effect Perception}

\Section{sec:DPW} outlined the use of sigmoid functions in identifying possible cognitive effect thresholds, nonlinearities, and listener preferences linked to distortion salience using DPWs. This technique significantly boosted quality prediction across all databases, as evidenced by comparing PEAQ-CSM and PEAQ-CSM+ performance figures in \Section{sec:validation}. \Figure{fig:OPTsfig} overlays the quality prediction effect of both systems in subjective/objective score scatter diagrams on an unseen, varied database, the MPEG-4 Enhanced Low Delay AAC Application Verification Tests -- ELD VT (A)\citep{ELDdatabase}. The improved linearity of system prediction is notable, particularly in mid to lower quality ranges where the optimized model's regression fit is closer to the $y=x$ line, signifying perfect prediction. The DPW modeling appears to enhance quality prediction where louder and multiple disturbances contribute to quality degradation.

\subsection{Quality Prediction Performance Generalization}
\label{sec:quality_prediction_generalization}
The results in \Section{sec:validation} show that, in contrast to other alternative mapping stages, the CSM mapping accurately predicts the quality of degraded signals in the lower quality range (i.e., USAC VT1 and ELD Technology Verification Tests -- ELD VT (T)) as well as in the higher quality range (ELD VT (A) and MPEG-H). The databases include subjective scores of a wide variety of audio signals and treatments from audio codecs, so the satisfactory performance of the model is expected to generalize to a wide range of use cases in the field of audio coding.

The superior performance of the CSM architecture is attributed to at least two key aspects. Firstly, the adoption of a multi-dimensional methodology is believed to be advantageous in effectively handling diverse types of signal degradation. The accepted notion that quality perception is inherently multi-dimensional, as highlighted by Letowski (1989) \cite{letowski1989sound}, reinforces our findings that methods like ViSQOL, PEMO-Q, and GPSMq, lacking distinct distortion metrics for different aspects of quality degradation, face challenges in addressing the diverse types of disturbances in the different databases. In contrast, the multi-dimensional approach of PEAQ-CSM+ is more effective in this respect. Compare, for example, the performance of the proposed method against ViSQOL in \Figure{fig:ViSQOLsfig}. Yet, relying solely on multiple distortion metrics does not explain the improved generalization, as the multi-dimensional method PEAQ DI doesn't outperform PEMO-Q significantly, even though the latter method does not employ multiple metrics for measuring degradation.

A second cause for improved performance is attributed to the CSM's unique approach to handling input variables, segregating DMs from CEMs as a consequence of the perceptually-motivated model architecture. In contrast to the other investigated general purpose ML algorithms, the CEMs are only allowed to interact with DMs by acting as adaptive weights, based on salience prediction (as opposed to allowing arbitrary interactions between input variables). In addition, the CEMs are not allowed to be mapped directly to the output layer. These restrictions effectively reduce model complexity in terms of number of parameters used, an important factor that causes overfitting. By adaptively weighting DMs using cognitive effects to predict distortion salience, the proposed system can measure quality degradations more accurately in a wider range of degraded signals than systems using general-purpose ML mapping stages that have been trained with the same data set (see e.g., \Figure{fig:ANNsfig}), and with larger datasets as in PEAQ's DI metric (see \Figure{fig:PEAQsfig}).

For the USAC VT 2 database, subjective score prediction is notably challenging, as seen from the weak correlation across all methods. Uniquely, this database primarily consists of stereo files coded at low bitrates, heavily using parametric stereo tools \cite{USACCodec}. These tools challenge quality measurement systems, since they do not maintain the original signal waveform and require advanced binaural models to gauge spatial auditory image degradations \cite{DelgadoICASSP}. Typical prominent spatial image distortions are associated with differences in the perceived localization of auditory sources, spatial image collapse and inaccurate envelopment sense reproduction \cite{flener2019, what_to_listen_for_2}. In our case, most analyzed systems only offer rudimentary stereo signal analyses (i.e., combining objective scores of separate channels). An exception is CombAQ, which incorporates a psychoacoustically validated binaural model. Although binaural models have shown efficacy in predicting spatial audio quality degradation \citep{seo2013perceptual, DelgadoICASSP}, CombAQ has not been trained with audio coding databases. We assume that CombAQ might show clear improvements if tuned and validated to signals distorted with audio coding processes.

\subsection{Performance on Application-Oriented Signals and Critical Signals for Parametric Codecs}

A key result is the proposed method's superior performance in databases that predominantly use parametric codecs, specifically the ELD VT databases. Most of the other tested methods, except for PEMO-Q PSMt (FB), did not perform as well under the same conditions. The parametric representations of signal waveforms stemming from non-waveform preserving codecs, often lead to noisy measurements in the corresponding DMs derived from internal representations, regardless of the number of smoothing stages, cochlear excitation models and loudness compression transformations applied (see i.e., \cite{Par2056_2019}).

The ELD VT databases introduce additional challenges for quality measurement due to the presence of critical signals such as mixed speech and music content, overlapping speakers, and background noise. These signal content selections are frequently encountered in teleconferencing scenarios. Cognitive metrics can enhance perceptual models, for instance, by using IM and PS to accurately measure distortions of competing sound percepts, and choosing the relevant DMs for speech or music quality perception using the speech probability metric. Yet, the use of cognitive effects alone does not necessarily result in improved quality prediction, as the results have demonstrated for general-purpose ML models assessed in \Section{sec:validation}.

\begin{figure}[htb]
  \begin{center}
      \resizebox{\columnwidth}{!}{
%
%
\begin{tikzpicture}

    \begin{axis}[%
        width=13.313cm,
        height=9cm,
        at={(0cm,0cm)},
        scale only axis,
        unbounded coords=jump,
        xmin=0.5,
        xmax=7.5,
        xtick={1,2,3,4,5,6,7},
        xticklabels={{LinDist/probSpeech ($Q_1$)},{NoiseLoudness/probSpeech ($Q_2$)},{EHS/probSpeech ($Q_3$)},{LinDist/EPN ($Q_4$)},{EHS/EPN ($Q_5$)},{NoiseLoudness/EPN/PDEV ($Q_6$)},{\hspace{-9.5cm} RmsModDiff ($Q_7$)}},
        xticklabel style={font=\normalsize, rotate=90, inner sep=0pt, anchor=north, xshift=6.3cm, yshift=-0.1cm},
        xlabel style={font=\color{white!15!black},font=\Large},
        xlabel={Quality Term},
        ymin=-3.56078031253979,
        ymax=38.1158709759501,
        ylabel style={font=\color{white!15!black},font=\Large},
        ylabel={Quality Score},
        yticklabel style={font=\Large},
        axis background/.style={fill=white},
        title style={font=\Large},
        title={Database: ELD VT (T)},
        xmajorgrids,
        ymajorgrids,
        legend style={font=\tiny},
        ]
    \addplot [color=black, dashed, forget plot, thick]
      table[row sep=crcr]{%
    1	10.2920552107395\\
    1	11.5351434404273\\
    };
    \addplot [color=black, dashed, forget plot, thick]
      table[row sep=crcr]{%
    2	9.60696667232806\\
    2	13.7314356639303\\
    };
    \addplot [color=black, dashed, forget plot, thick]
      table[row sep=crcr]{%
    3	2.77079257430332\\
    3	4.03198407159623\\
    };
    \addplot [color=black, dashed, forget plot, thick]
      table[row sep=crcr]{%
    4	6.30503134126333\\
    4	6.73460684828942\\
    };
    \addplot [color=black, dashed, forget plot, thick]
      table[row sep=crcr]{%
    5	3.13006913090532\\
    5	5.09752805667164\\
    };
    \addplot [color=black, dashed, forget plot, thick]
      table[row sep=crcr]{%
    6	-1.81080099876257e-05\\
    6	-8.28825132695749e-06\\
    };
    \addplot [color=black, dashed, forget plot, thick]
      table[row sep=crcr]{%
    7	30.3421970351209\\
    7	36.2214777355642\\
    };
    \addplot [color=black, dashed, forget plot, thick]
      table[row sep=crcr]{%
    1	2.57403974660339\\
    1	7.09990403611358\\
    };
    \addplot [color=black, dashed, forget plot, thick]
      table[row sep=crcr]{%
    2	0.831084881812355\\
    2	5.02542472792608\\
    };
    \addplot [color=black, dashed, forget plot, thick]
      table[row sep=crcr]{%
    3	0.632424645444497\\
    3	1.38718009548981\\
    };
    \addplot [color=black, dashed, forget plot, thick]
      table[row sep=crcr]{%
    4	1.02714770985831\\
    4	4.179729912809\\
    };
    \addplot [color=black, dashed, forget plot, thick]
      table[row sep=crcr]{%
    5	0.984180366692023\\
    5	1.46155306959041\\
    };
    \addplot [color=black, dashed, forget plot, thick]
      table[row sep=crcr]{%
    6	-0.000115703151381398\\
    6	-7.9794608109161e-05\\
    };
    \addplot [color=black, dashed, forget plot, thick]
      table[row sep=crcr]{%
    7	12.0412059307167\\
    7	22.1062053998234\\
    };
    \addplot [color=black, forget plot, thick]
      table[row sep=crcr]{%
    0.875	11.5351434404273\\
    1.125	11.5351434404273\\
    };
    \addplot [color=black, forget plot, thick]
      table[row sep=crcr]{%
    1.875	13.7314356639303\\
    2.125	13.7314356639303\\
    };
    \addplot [color=black, forget plot, thick]
      table[row sep=crcr]{%
    2.875	4.03198407159623\\
    3.125	4.03198407159623\\
    };
    \addplot [color=black, forget plot, thick]
      table[row sep=crcr]{%
    3.875	6.73460684828942\\
    4.125	6.73460684828942\\
    };
    \addplot [color=black, forget plot, thick]
      table[row sep=crcr]{%
    4.875	5.09752805667164\\
    5.125	5.09752805667164\\
    };
    \addplot [color=black, forget plot, thick]
      table[row sep=crcr]{%
    5.875	-8.28825132695749e-06\\
    6.125	-8.28825132695749e-06\\
    };
    \addplot [color=black, forget plot, thick]
      table[row sep=crcr]{%
    6.875	36.2214777355642\\
    7.125	36.2214777355642\\
    };
    \addplot [color=black, forget plot, thick]
      table[row sep=crcr]{%
    0.875	2.57403974660339\\
    1.125	2.57403974660339\\
    };
    \addplot [color=black, forget plot, thick]
      table[row sep=crcr]{%
    1.875	0.831084881812355\\
    2.125	0.831084881812355\\
    };
    \addplot [color=black, forget plot, thick]
      table[row sep=crcr]{%
    2.875	0.632424645444497\\
    3.125	0.632424645444497\\
    };
    \addplot [color=black, forget plot, thick]
      table[row sep=crcr]{%
    3.875	1.02714770985831\\
    4.125	1.02714770985831\\
    };
    \addplot [color=black, forget plot, thick]
      table[row sep=crcr]{%
    4.875	0.984180366692023\\
    5.125	0.984180366692023\\
    };
    \addplot [color=black, forget plot, thick]
      table[row sep=crcr]{%
    5.875	-0.000115703151381398\\
    6.125	-0.000115703151381398\\
    };
    \addplot [color=black, forget plot, thick]
      table[row sep=crcr]{%
    6.875	12.0412059307167\\
    7.125	12.0412059307167\\
    };
    \addplot [color=black, forget plot, thick]
      table[row sep=crcr]{%
    0.75	7.09990403611358\\
    0.75	10.2920552107395\\
    1.25	10.2920552107395\\
    1.25	7.09990403611358\\
    0.75	7.09990403611358\\
    };
    \addplot [color=black, forget plot, thick]
      table[row sep=crcr]{%
    1.75	5.02542472792608\\
    1.75	9.60696667232806\\
    2.25	9.60696667232806\\
    2.25	5.02542472792608\\
    1.75	5.02542472792608\\
    };
    \addplot [color=black, forget plot, thick]
      table[row sep=crcr]{%
    2.75	1.38718009548981\\
    2.75	2.77079257430332\\
    3.25	2.77079257430332\\
    3.25	1.38718009548981\\
    2.75	1.38718009548981\\
    };
    \addplot [color=black, forget plot, thick]
      table[row sep=crcr]{%
    3.75	4.179729912809\\
    3.75	6.30503134126333\\
    4.25	6.30503134126333\\
    4.25	4.179729912809\\
    3.75	4.179729912809\\
    };
    \addplot [color=black, forget plot, thick]
      table[row sep=crcr]{%
    4.75	1.46155306959041\\
    4.75	3.13006913090532\\
    5.25	3.13006913090532\\
    5.25	1.46155306959041\\
    4.75	1.46155306959041\\
    };
    \addplot [color=black, forget plot, thick]
      table[row sep=crcr]{%
    5.75	-7.9794608109161e-05\\
    5.75	-1.81080099876257e-05\\
    6.25	-1.81080099876257e-05\\
    6.25	-7.9794608109161e-05\\
    5.75	-7.9794608109161e-05\\
    };
    \addplot [color=black, forget plot, thick]
      table[row sep=crcr]{%
    6.75	22.1062053998234\\
    6.75	30.3421970351209\\
    7.25	30.3421970351209\\
    7.25	22.1062053998234\\
    6.75	22.1062053998234\\
    };
    \addplot [color=gray, forget plot, thick]
      table[row sep=crcr]{%
    0.75	8.55226062379102\\
    1.25	8.55226062379102\\
    };
    \addplot [color=gray, forget plot, thick]
      table[row sep=crcr]{%
    1.75	6.41656970140193\\
    2.25	6.41656970140193\\
    };
    \addplot [color=gray, forget plot, thick]
      table[row sep=crcr]{%
    2.75	1.99384191763878\\
    3.25	1.99384191763878\\
    };
    \addplot [color=gray, forget plot, thick]
      table[row sep=crcr]{%
    3.75	5.05037705842216\\
    4.25	5.05037705842216\\
    };
    \addplot [color=gray, forget plot, thick]
      table[row sep=crcr]{%
    4.75	2.71145848949258\\
    5.25	2.71145848949258\\
    };
    \addplot [color=gray, forget plot, thick]
      table[row sep=crcr]{%
    5.75	-2.59205555891053e-05\\
    6.25	-2.59205555891053e-05\\
    };
    \addplot [color=gray, forget plot, thick]
      table[row sep=crcr]{%
    6.75	27.3612287979407\\
    7.25	27.3612287979407\\
    };
    \addplot [color=black, only marks, mark=+, mark options={solid, draw=gray}, forget plot, thick]
      table[row sep=crcr]{%
    nan	nan\\
    };
    \addplot [color=black, only marks, mark=+, mark options={solid, draw=gray}, forget plot, thick]
      table[row sep=crcr]{%
    nan	nan\\
    };
    \addplot [color=black, only marks, mark=+, mark options={solid, draw=gray}, forget plot, thick]
      table[row sep=crcr]{%
    nan	nan\\
    };
    \addplot [color=black, only marks, mark=+, mark options={solid, draw=gray}, forget plot, thick]
      table[row sep=crcr]{%
    nan	nan\\
    };
    \addplot [color=black, only marks, mark=+, mark options={solid, draw=gray}, forget plot, thick]
      table[row sep=crcr]{%
    5	5.85877809655987\\
    5	6.40940670860405\\
    };
    \addplot [color=black, only marks, mark=+, mark options={solid, draw=gray}, forget plot, thick]
      table[row sep=crcr]{%
    6	-1.66638707215389\\
    6	-1.42507029190271\\
    6	-0.189082933911532\\
    6	-0.124686100538767\\
    6	-0.0544200696909765\\
    6	-0.00577295464451976\\
    6	-0.00121736752564696\\
    6	-0.00104417002434811\\
    6	-0.000345665743435831\\
    6	-0.000255476836064954\\
    6	-0.000235024996163819\\
    6	-0.000208424865901186\\
    };
    \addplot [color=black, only marks, mark=+, mark options={solid, draw=gray}, forget plot, thick]
      table[row sep=crcr]{%
    nan	nan\\
    };
    \end{axis}
    \end{tikzpicture}%
    
      }
  \end{center}
  \caption{Boxplot depicting the values of the objective quality terms of \Table{tab:importanceOptim} evaluated on the ELD VT (T) database, for the proposed PEAQ-CSM+ method.}
  \label{fig:CH_timbre__boxplot_CSM_ELD_SBR}
  \end{figure}
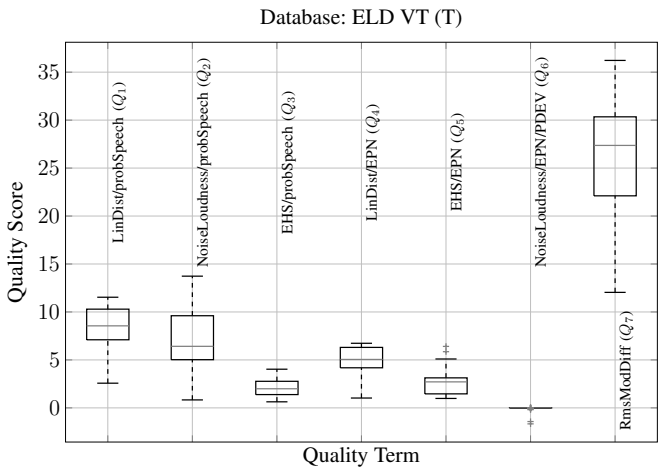

  \begin{figure}[htb]
  \begin{center}
      \resizebox{\columnwidth}{!}{
        \input{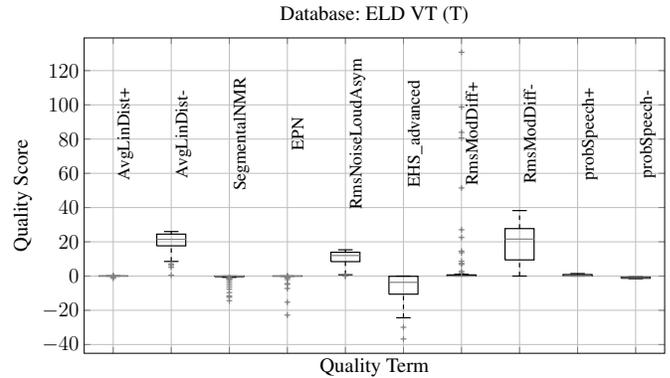}
      }
  \end{center}
  \caption{Boxplot depicting the values of the objective quality terms of a MARS model trained with the same data and input variables as the proposed PEAQ-CSM+ method, evaluated on the ELD VT (T) database.}
  \label{fig:CH_timbre__boxplot_PEAQCEMS_ELD_SBR}
\end{figure}

The superior performance of the CSM can apparently be attributed to its perceptually-inspired architecture that employs cognitive effects to stabilize the values of different quality terms in \Equation{eq:sum_Q}, which all contribute to the total objective score. In contrast, a general-purpose ML algorithm does not consider cognitive effects in this manner. Comparing the distribution of the quality terms of the proposed system using the CSM model (\Figure{fig:CH_timbre__boxplot_CSM_ELD_SBR}) and one of the evaluated general-purpose ML models (MARS), trained on the same dataset (\Figure{fig:CH_timbre__boxplot_PEAQCEMS_ELD_SBR}), reveals a noticeable difference. The quality term values of the MARS model do not contribute as evenly as the terms in the CSM model: many of the value distributions are centered around 0 with various outliers with negative quality score contributions. Also note that some cognitive effects (EPN, probSpeech) directly contribute to the quality score in the MARS model as independent quality terms. Ideally, however, CEMs should act as modulators of the DMs in the form of multiplying factors, as is the case of the CSM, rather than as direct contributors to the final objective quality score in the form of independent quality terms.

The importance of the contributions from different CSM terms is determined by the regression coefficient weights shown in \Table{tab:importanceOptim}. A great portion of the objective assessment score is influenced by $Q_7$, related to modulation distortions, which also has the largest value spread. The significance of a modulation metric for quality measurement of parametrically coded signals is further supported by the work of Van de Par et al., in \citep{Par2056_2019}. The effective performance of PEMO-Q PSMt (FB) on ELD VT also supports this result, as this quality system measures modulation distortions in coded signals by applying metrics from a built-in modulation filter bank. Modulation distortion metrics have also proven to be meaningful for the quality estimation of separated audio source signals \cite{2fKastner}. However, the results also suggest that solely relying on a modulation metric is insufficient to assure satisfactory performance across all other evaluated databases. In the CSM, other stabilized quality terms associated with additional distortion types are used, with the quality metrics maintaining the correct polarity, except for $Q_6$, which could possibly indicate scale border effects in cases of significant disturbances as explained previously.

\subsection{Generalization to Other Application Domains}
The proposed model effectively predicts quality degradation across a wide range of signals and distortions related to audio coding. It also demonstrates satisfactory performance on the BSS database SASSEC, where most of the other analyzed state-of-the-art systems struggle. As discussed in \cite{TorcoliApplication}, systems optimized for audio coding degradation may not reliably predict degradations in other domains. Despite being trained and tuned with audio coding distortion databases, PEAQ-CSM also performs well in other application domains due to the diversity of its training data. However, data diversity alone is not sufficient; a robust perceptual model is also crucial for analyzing and distilling the hidden relationships in the data, such as the relationship between distortion salience and signal type. The results presented in \Figure{tab:heatmap_overall_R} for the SASSEC database further support these observations, and they are also consistent with findings in \citep{TorcoliApplication}, where PEMO-Q also performs well in BSS. We plan to validate PEAQ-CSM on further application domains in the future, provided that sufficiently large validation data is available. 

\subsection{Computational Complexity}

The results in \Table{tab:table_complexity} show that the measured execution time is expectedly dependent on the number of model parameters and the chosen architecture. Most models have a similar order of magnitude in parameter count, except for the KSOM variant. Initial experiments showed that even a slight reduction in KSOM's parameter count significantly degraded prediction performance because a high parameter count is needed for achieving the MUSHRA scale's granularity on the output classifier. The inference time for PEAQ-CSM is comparable to that of MARS due to PEAQ-CSM's use of similarly complex spline-based basis functions, and considerably less than the calculation time for the baseline ANN mapping. However, the increased generalization power gained from using sigmoid functions for weighing the CEMs results in a 22\% increase in calculation time. Nonetheless, the inference time for both variants of the proposed model is negligible compared to the total calculation time in the perceptual model: The MATLAB implementation took about 4 seconds to calculate DMs and CEMs. For reference, the C implementation of PEAQ presented in \cite{holters2015gstpeaq} took about 0.9 seconds for the similar task of calculating the MOVs while the speech classification task took around 0.25 seconds. Similar calculation times can be expected for a faster implementation of PEAQ-CSM.

\subsection{Limitations}
Firstly, objective metrics often show varying subjective quality prediction performance across different signal types. For instance, the work in \citep{ThiedePHD} highlighted PEAQ's limitations with male speech items. The incorporation of a speech classifier in the proposed model of this work improved quality prediction relative to PEAQ's baseline, aligning with the suggestions in \citep{ThiedePHD} to use signal recognition processes to improve speech quality assessment. However, some issues persist, as highlighted in \Section{sec:csm_model_parameters}, particularly with the model predicting the quality of low-rated speech signals on the MUSHRA scale. 

Furthermore, a performance analysis of PEAQ-CSM+ in the USAC VT1 database revealed that the predictions sometimes fell outside the subjective score confidence intervals, particularly for three music signals and one speech item with music sound effects. Specifically, the system overestimated quality degradation for items processed by HE-AACv2 at 24 kbps in mono operation. This codec is known for its extensive use of parametric, non-waveform-preserving tools, which are generally known to represent a challenge for perceptual quality measurement. In \cite{delgado2023improved}, we thus proposed a refined IM model that improved quality prediction in this particular case, by interacting with the EHS DM. Nevertheless, in an overall view, when the predictions of the presented model were averaged over all signals in USAC VT1, the predictions fell within the confidence interval of the pooled overall subjective scores for the HE-AACv2 condition, and no rank inversions were observed for any of the remaining codecs. 

Finally, as highlighted in \Section{sec:quality_prediction_generalization}, the proposed model has clear limitations in predicting the quality for signals with significant stereo field distortions in the USAC VT2 database, as it completely lacks a spatial perceptual model, and is therefore blind to these type of degradations. 

\subsection{Future Work}

The suitability of the proposed model in application domains other than audio coding and BSS remains to be studied. While the proposed model was mostly validated with MUSHRA-type listening tests and with one BS.1116 database, future work will include validating and extending the scheme to different experimental designs like P.800 tests \cite{ITUP800}. Such test methodologies are commonly used for speech communication codecs, where issues like frame drop-outs and decreased speech intelligibility can play a significant role in quality degradation. Recently proposed metrics for speech quality assessment incorporate multiple quality degradation predictors \cite{mittag2021nisqa}. Applying similar analyses of distortion salience and cognitive effects could thus enhance objective quality assessment in this case. Additionally, further investigations may include the evaluation of the model’s performance without an original undisturbed reference (non-intrusive measurement). Finally, and most importantly, future work will extend the system with a binaural perceptual model to enable adequate assessment of spatial distortions.

\section{Conclusions}
This study introduced a perceptually-motivated ML architecture as an alternative cognitive model for the standardized audio quality measurement system PEAQ. The proposed model explicitly considers interactions between cognitive effects and distortion salience. Results indicate that this system shows improved generalization prediction performance compared to other ML algorithms trained on the same dataset and other state-of-the-art quality metrics. Crucially, the model's predictions align well with psychoacoustic results, highlighting its effectiveness in capturing perceptual nuances. Additionally, the flexibility of the proposed architecture and training procedure makes it a versatile framework that can be extended to include additional psychoacoustic effects in various application scenarios beyond PEAQ's original scope.

\section*{Acknowledgments}

The authors would like to thank Thomas Sporer for his valuable feedback. Additional thanks to Martin Müller, Markus Schnell, Stefan Emmert, Aalok Gupta and Madhurya Kumar Dutta for their support.



\begin{thebibliography}{10}
\providecommand{\url}[1]{#1}
\csname url@samestyle\endcsname
\providecommand{\newblock}{\relax}
\providecommand{\bibinfo}[2]{#2}
\providecommand{\BIBentrySTDinterwordspacing}{\spaceskip=0pt\relax}
\providecommand{\BIBentryALTinterwordstretchfactor}{4}
\providecommand{\BIBentryALTinterwordspacing}{\spaceskip=\fontdimen2\font plus
\BIBentryALTinterwordstretchfactor\fontdimen3\font minus
  \fontdimen4\font\relax}
\providecommand{\BIBforeignlanguage}[2]{{%
\expandafter\ifx\csname l@#1\endcsname\relax
\typeout{** WARNING: IEEEtran.bst: No hyphenation pattern has been}%
\typeout{** loaded for the language `#1'. Using the pattern for}%
\typeout{** the default language instead.}%
\else
\language=\csname l@#1\endcsname
\fi
#2}}
\providecommand{\BIBdecl}{\relax}
\BIBdecl

\bibitem{brandenburg1994iso}
K.~Brandenburg and G.~Stoll, ``{ISO/MPEG-1} audio: A generic standard for
  coding of high-quality digital audio,'' \emph{Journal of the Audio
  Engineering Society}, vol.~42, no.~10, pp. 780--792, 1994.

\bibitem{OverviewMediaCompression2021}
C.~Timmerer, M.~Wien, L.~Yu, and A.~Reibman, ``Special issue on open media
  compression: Overview, design criteria, and outlook on emerging standards,''
  \emph{Proceedings of the IEEE}, vol. 109, no.~9, pp. 1423--1434, 2021.

\bibitem{RixTaslp}
A.~W. {Rix}, J.~G. {Beerends}, D.~. {Kim}, P.~{Kroon}, and O.~{Ghitza},
  ``Objective assessment of speech and audio quality—technology and
  applications,'' \emph{IEEE Transactions on Audio, Speech, and Language
  Processing}, vol.~14, no.~6, pp. 1890--1901, 2006.

\bibitem{PEAQ}
{\relax ITU-R Rec. BS.1387}, \emph{Method for objective measurements of
  perceived audio quality}, Geneva, Switzerland, 2001.

\bibitem{POLQAcite}
{\relax ITU-T Rec. P.863}, \emph{Perceptual Objective Listening Quality
  Assessment}, Geneva, Switzerland, 2014.

\bibitem{ViSQOLAudio}
C.~Sloan, N.~Harte, D.~Kelly, A.~C. Kokaram, and A.~Hines, ``Objective
  assessment of perceptual audio quality using {ViSQOLAudio},'' \emph{IEEE
  Transactions on Broadcasting}, vol.~PP, no.~99, pp. 1--13, 2017.

\bibitem{PEMOQ}
R.~{Huber} and B.~{Kollmeier}, ``{PEMO-Q}—a new method for objective audio
  quality assessment using a model of auditory perception,'' \emph{IEEE
  Transactions on Audio, Speech, and Language Processing}, vol.~14, no.~6, pp.
  1902--1911, Nov 2006.

\bibitem{defossez2022high}
A.~Défossez, J.~Copet, G.~Synnaeve, and Y.~Adi, ``High fidelity neural audio
  compression,'' 2022.

\bibitem{FingscheidtPESQ}
Z.~Xu, M.~Strake, and T.~Fingscheidt, ``Deep noise suppression maximizing
  non-differentiable {PESQ} mediated by a non-intrusive {PESQNet},''
  \emph{IEEE/ACM Transactions on Audio, Speech, and Language Processing},
  vol.~30, pp. 1572--1585, 2022.

\bibitem{TorcoliApplication}
M.~Torcoli, T.~Kastner, and J.~Herre, ``Objective measures of perceptual audio
  quality reviewed: An evaluation of their application domain dependence,''
  \emph{IEEE/ACM Transactions on Audio, Speech, and Language Processing},
  vol.~29, pp. 1530--1541, 2021.

\bibitem{PESQ}
{ITU-R Rec. P.862}, \emph{{Perceptual evaluation of speech quality (PESQ)}},
  Geneva, Switzerland, Dec. 2001.

\bibitem{ThiedePEAQ}
T.~Thiede, W.~C. Treurniet, R.~Bitto, C.~Schmidmer, T.~Sporer, J.~G. Beerends,
  and C.~Colomes, ``{PEAQ} - the {ITU} standard for objective measurement of
  perceived audio quality,'' \emph{J. Audio Eng. Soc.}, vol.~48, no. 1/2, pp.
  3--29, January/February 2000.

\bibitem{RixPhD}
A.~Rix, ``Perceptual techniques in audio quality assessment,'' Ph.D.
  dissertation, University of Edinburgh, 2003.

\bibitem{SoniDeepSpeechQuality}
M.~H. Soni and H.~A. Patil, ``Novel deep autoencoder features for non-intrusive
  speech quality assessment,'' in \emph{2016 24th European Signal Processing
  Conference (EUSIPCO)}, 2016, pp. 2315--2319.

\bibitem{NetFlix}
C.-W. Wu, P.~A. Williams, and W.~Wolcott, ``A multitask teacher-student
  framework for perceptual audio quality assessment,'' in \emph{2021 29th
  European Signal Processing Conference (EUSIPCO)}, 2021, pp. 396--400.

\bibitem{manocha2020differentiable}
P.~Manocha, A.~Finkelstein, R.~Zhang, N.~J. Bryan, G.~J. Mysore, and Z.~Jin,
  ``A differentiable perceptual audio metric learned from just noticeable
  differences,'' in \emph{Interspeech}, Oct. 2020.

\bibitem{jiang2023generative}
G.~Jiang, L.~Villemoes, and A.~Biswas, ``Generative machine listener,'' in
  \emph{Audio Engineering Society Convention 155}, Oct 2023.

\bibitem{delgado2022data}
P.~M. Delgado and J.~Herre, ``A data-driven cognitive salience model for
  objective perceptual audio quality assessment,'' in \emph{ICASSP 2022-2022
  IEEE International Conference on Acoustics, Speech and Signal Processing
  (ICASSP)}.\hskip 1em plus 0.5em minus 0.4em\relax IEEE, 2022, pp. 986--990.

\bibitem{bregman1994auditory}
A.~S. Bregman, \emph{Auditory scene analysis: The perceptual organization of
  sound}.\hskip 1em plus 0.5em minus 0.4em\relax MIT press, 1994.

\bibitem{MUSHRA}
{\relax ITU-R Rec. BS.1534}, \emph{{Method for the subjective assessment of
  intermediate quality levels of coding systems}}, Geneva, Switzerland, 2015.

\bibitem{BS1116}
{\relax ITU-R Rec. BS.1116}, \emph{{Methods for the subjective assessment of
  small impairments in audio systems}}, Geneva, Switzerland, 2015.

\bibitem{letowski1989sound}
\BIBentryALTinterwordspacing
T.~Letowski, ``Sound quality assessment: Concepts and criteria,'' in
  \emph{Audio Engineering Society Convention 87}, New York, Oct 1989. [Online].
  Available: \url{http://www.aes.org/e-lib/browse.cfm?elib=5869}
\BIBentrySTDinterwordspacing

\bibitem{flener2017assessment}
\BIBentryALTinterwordspacing
J.-H. Fleßner, R.~Huber, and S.~D. Ewert, ``Assessment and prediction of
  binaural aspects of audio quality,'' \emph{J. Audio Eng. Soc}, vol.~65,
  no.~11, pp. 929--942, 2017. [Online]. Available:
  \url{http://www.aes.org/e-lib/browse.cfm?elib=19361}
\BIBentrySTDinterwordspacing

\bibitem{DelgadoPEAQ}
P.~M. {Delgado} and J.~{Herre}, ``Can we still use {PEAQ}? {A} performance
  analysis of the {ITU} standard for the objective assessment of perceived
  audio quality,'' in \emph{2020 Twelfth International Conference on Quality of
  Multimedia Experience (QoMEX)}, 2020, pp. 1--6.

\bibitem{barbedo2005a}
\BIBentryALTinterwordspacing
J.~G.~A. Barbedo and A.~Lopes, ``A new cognitive model for objective assessment
  of audio quality,'' \emph{J. Audio Eng. Soc}, vol.~53, no. 1/2, pp. 22--31,
  2005. [Online]. Available:
  \url{http://www.aes.org/e-lib/browse.cfm?elib=13386}
\BIBentrySTDinterwordspacing

\bibitem{SpeechEvaluationMultipleForeground}
\BIBentryALTinterwordspacing
D.~Sen and W.~Lu, ``Objective evaluation of speech signal quality by the
  prediction of multiple foreground diagnostic acceptability measure
  attributes,'' \emph{The Journal of the Acoustical Society of America}, vol.
  131, no.~5, pp. 4087--4103, 2012. [Online]. Available:
  \url{https://doi.org/10.1121/1.3699262}
\BIBentrySTDinterwordspacing

\bibitem{MATLAB}
{The MathWorks}, \emph{version 9.14.0 (R2023a)}.\hskip 1em plus 0.5em minus
  0.4em\relax Natick, Massachusetts: The MathWorks Inc., 2023.

\bibitem{holters2015gstpeaq}
M.~Holters and U.~Z{\"o}lzer, ``{GstPEAQ}--an open source implementation of the
  {PEAQ} algorithm,'' in \emph{Proc. 18th Int. Conf. Digital Audio Effects
  (DAFx), Trondheim, Norway}, 2015.

\bibitem{beerends1992a}
\BIBentryALTinterwordspacing
J.~G. Beerends and J.~A. Stemerdink, ``A perceptual audio quality measure based
  on a psychoacoustic sound representation,'' \emph{J. Audio Eng. Soc},
  vol.~40, no.~12, pp. 963--978, 1992. [Online]. Available:
  \url{http://www.aes.org/e-lib/browse.cfm?elib=7019}
\BIBentrySTDinterwordspacing

\bibitem{Zwickerbook}
E.~Zwicker and H.~Fastl, \emph{Psychoacoustics. Facts and Models}.\hskip 1em
  plus 0.5em minus 0.4em\relax Springer, 1999.

\bibitem{ThiedePHD}
T.~Thiede, ``Perceptual audio quality assessment using a non-linear filter
  bank,'' Ph.D. dissertation, Fachbereich Elektrotechnik, Technische
  Universit{ä}t Berlin, 1999.

\bibitem{USACSpeechClas}
\BIBentryALTinterwordspacing
G.~Fuchs, ``A robust speech/music discriminator for switched audio coding,'' in
  \emph{23rd European Signal Processing Conference, {EUSIPCO} 2015, Nice,
  France, August 31 - September 4, 2015}.\hskip 1em plus 0.5em minus
  0.4em\relax {IEEE}, 2015, pp. 569--573. [Online]. Available:
  \url{https://doi.org/10.1109/EUSIPCO.2015.7362447}
\BIBentrySTDinterwordspacing

\bibitem{what_to_listen_for_2}
\BIBentryALTinterwordspacing
S.~Dick and {AES Technical Committee on Coding of Audio Signals (TC-CAS)},
  ``{Perceptual Audio Codecs - What to Listen For, Web Edition},'' in
  \emph{Audio Engineering Society}, May 2021. [Online]. Available:
  \url{https://aes2.org/resources/audio-topics/audio_coding/perceptual-audio-codecs/}
\BIBentrySTDinterwordspacing

\bibitem{beerends1996the}
\BIBentryALTinterwordspacing
J.~G. Beerends, W.~A.~C. van~den Brink, and B.~Rodger, ``The role of
  informational masking and perceptual streaming in the measurement of music
  codec quality,'' in \emph{Audio Engineering Society Convention 100},
  Copenhagen, May 1996. [Online]. Available:
  \url{http://www.aes.org/e-lib/browse.cfm?elib=7596}
\BIBentrySTDinterwordspacing

\bibitem{SpeechVsMusicQuality}
\BIBentryALTinterwordspacing
R.~Huber, S.~Rählmann, T.~Bisitz, M.~Meis, S.~Steinhauser, and H.~Meister,
  ``Influence of working memory and attention on sound-quality ratings,''
  \emph{The Journal of the Acoustical Society of America}, vol. 145, no.~3, pp.
  1283--1292, 2019. [Online]. Available:
  \url{https://doi.org/10.1121/1.5092808}
\BIBentrySTDinterwordspacing

\bibitem{Jekabsons_areslab}
G.~Jekabsons, ``{ARESLab}: Adaptive regression splines toolbox for {MATLAB}.
  http://www.cs.rtu. lv/jekabsons/regression.html,'' 2019.

\bibitem{PsychometricCitation}
\BIBentryALTinterwordspacing
S.~Buus, C.~R. Mason, and M.~Florentine, ``Psychometric functions for level
  discrimination,'' \emph{The Journal of the Acoustical Society of America},
  vol.~82, no.~S1, pp. S25--S25, 1987. [Online]. Available:
  \url{https://doi.org/10.1121/1.2024720}
\BIBentrySTDinterwordspacing

\bibitem{hastie2009elements}
T.~Hastie, R.~Tibshirani, and J.~Friedman, \emph{The Elements of Statistical
  Learning: Data Mining, Inference, and Prediction}, ser. Springer series in
  statistics.\hskip 1em plus 0.5em minus 0.4em\relax Springer, 2009.

\bibitem{MPEGHdatabase}
{\relax ISO/IEC JTC1/SC29/WG11}, ``Submission and evaluation procedures for
  {3D} audio {N13633},'' International Organisation for Standardisation, Tech.
  Rep., 2013.

\bibitem{ITUDB45}
{ITU-R BS.1387-1}, ``{Progress report towards revision of recommendation ITU-R
  BS.1387-1 (Annex 13 to Document 6C/415-E)},'' Geneva, Switzerland, Nov. 2010.

\bibitem{ELDdatabase}
{\relax ISO/IEC JTC1/SC29/WG11}, ``Report on the verification test of {MPEG-4
  Enhanced Low Delay AAC N10032},'' International Organisation for
  Standardisation, Hannover, Germany, Tech. Rep., 2008.

\bibitem{USACdatabase}
------, ``{USAC} verification test report {N12232},'' International
  Organisation for Standardisation, Tech. Rep., 2011.

\bibitem{dick2017generation}
\BIBentryALTinterwordspacing
S.~Dick, N.~Schinkel-Bielefeld, and S.~Disch, ``Generation and evaluation of
  isolated audio coding artifacts,'' in \emph{Audio Engineering Society
  Convention 143}, New York, Oct 2017. [Online]. Available:
  \url{http://www.aes.org/e-lib/browse.cfm?elib=19206}
\BIBentrySTDinterwordspacing

\bibitem{SEBASS}
T.~Kastner and J.~Herre, ``The sebass-db: A consolidated public data base of
  listening test results for perceptual evaluation of bss quality measures,''
  in \emph{2022 International Workshop on Acoustic Signal Enhancement
  (IWAENC)}, 2022, pp. 1--5.

\bibitem{vincent2007first}
E.~Vincent, H.~Sawada, P.~Bofill, S.~Makino, and J.~P. Rosca, ``First stereo
  audio source separation evaluation campaign: data, algorithms and results,''
  in \emph{International Conference on Independent Component Analysis and
  Signal Separation}.\hskip 1em plus 0.5em minus 0.4em\relax Springer, 2007,
  pp. 552--559.

\bibitem{EvalObjective}
{\relax ITU-T Rec. P.1401}, \emph{Methods, metrics and procedures for
  statistical evaluation, qualification and comparison of objective quality
  prediction models}, Geneva, Switzerland, 2012.

\bibitem{RevisionBS1387}
ITU-R, ``Workplan towards draft revision of recommendation {ITU-R BS.1387-1},''
  Annex 12 to Working Party 6C Chairman's Report. International
  Telecommunication Union, Tech. Rep., 2009.

\bibitem{Flessner}
J.~H. Flessner, S.~D. Ewert, B.~Kollmeier, and R.~Huber, ``Quality assessment
  of multi-channel audio processing schemes based on a binaural auditory
  model,'' in \emph{2014 IEEE International Conference on Acoustics, Speech and
  Signal Processing (ICASSP)}, May 2014, pp. 1340--1344.

\bibitem{silzle2009vision}
\BIBentryALTinterwordspacing
A.~Silzle, S.~Geyersberger, G.~Brohasga, D.~Weninger, and M.~Leistner, ``Vision
  and technique behind the new studios and listening rooms of the {Fraunhofer
  IIS} audio laboratory,'' in \emph{Audio Engineering Society Convention 126},
  Munich, May 2009. [Online]. Available:
  \url{http://www.aes.org/e-lib/browse.cfm?elib=14868}
\BIBentrySTDinterwordspacing

\bibitem{CombAQGit}
T.~{Biberger} and J.~{Fleßner}, ``{MATLAB implementation of the Combined Audio
  Quality Model},''
  \url{https://gitlab.uni-oldenburg.de/kuxo2262/combinedaudioqualitymodel},
  2019, accessed: 2022-07-07.

\bibitem{flener2019}
J.~{Fleßner}, T.~{Biberger}, and S.~D. {Ewert}, ``Subjective and objective
  assessment of monaural and binaural aspects of audio quality,''
  \emph{IEEE/ACM Transactions on Audio, Speech, and Language Processing},
  vol.~27, no.~7, pp. 1112--1125, 2019.

\bibitem{biberger2018objective}
T.~Biberger, J.-H. Fle{\ss}ner, R.~Huber, and S.~D. Ewert, ``An objective audio
  quality measure based on power and envelope power cues,'' \emph{Journal of
  the Audio Engineering Society}, vol.~66, no. 7/8, pp. 578--593, 2018.

\bibitem{emiya2010peass}
V.~Emiya, E.~Vincent, N.~Harlander, and V.~Hohmann, ``The {PEASS}
  toolkit-perceptual evaluation methods for audio source separation,'' in
  \emph{9th Int. Conf. on Latent Variable Analysis and Signal Separation},
  2010.

\bibitem{Visqol_soft}
A.~Hines, E.~Gillen, D.~Kelly, J.~Skoglund, A.~Kokaram, and N.~Harte, ``{ViSQOL
  Audio} {MATLAB} implementation. http://www.sigmedia.tv/tools,'' Accessed
  2019.

\bibitem{ArtifactsinPAC}
C.-M. Liu, H.-W. Hsu, and W.-C. Lee, ``Compression artifacts in perceptual
  audio coding,'' \emph{IEEE Transactions on Audio, Speech, and Language
  Processing}, vol.~16, no.~4, pp. 681--695, 2008.

\bibitem{HarmonicStructureGrouping}
B.~Roberts and J.~M. Brunstrom, ``Spectral pattern, harmonic relations, and the
  perceptual grouping of low-numbered components,'' \emph{The Journal of the
  Acoustical Society of America}, vol. 114, no.~4, pp. 2118--2134, 2003.

\bibitem{herre_1996_tns}
J.~Herre and D.~Johnston, ``Enhancing the performance of perceptual audio
  coders by using {T}emporal {N}oise {S}haping ({TNS}),'' in \emph{101st AES
  Convention}, Los Angeles, 1996, preprint 4384.

\bibitem{USACCodec}
\BIBentryALTinterwordspacing
M.~Neuendorf, M.~Multrus, N.~Rettelbach, G.~Fuchs, J.~Robilliard, J.~Lecomte,
  S.~Wilde, S.~Bayer, S.~Disch, C.~Helmrich, R.~Lefebvre, P.~Gournay,
  B.~Bessette, J.~Lapierre, K.~Kjörling, H.~Purnhagen, L.~Villemoes, W.~Oomen,
  E.~Schuijers, K.~Kikuiri, T.~Chinen, T.~Norimatsu, K.~S. Chong, E.~Oh,
  M.~Kim, S.~Quackenbush, and B.~Grill, ``The {ISO/MPEG} unified speech and
  audio coding standard—consistent high quality for all content types and at
  all bit rates,'' \emph{J. Audio Eng. Soc}, vol.~61, no.~12, pp. 956--977,
  2013. [Online]. Available:
  \url{http://www.aes.org/e-lib/browse.cfm?elib=17074}
\BIBentrySTDinterwordspacing

\bibitem{DelgadoICASSP}
P.~M. {Delgado} and J.~{Herre}, ``Objective assessment of spatial audio quality
  using directional loudness maps,'' in \emph{ICASSP 2019 - 2019 IEEE
  International Conference on Acoustics, Speech and Signal Processing
  (ICASSP)}, 2019.

\bibitem{seo2013perceptual}
\BIBentryALTinterwordspacing
J.-H. Seo, S.~B. Chon, K.-M. Sung, and I.~Choi, ``Perceptual objective quality
  evaluation method for high quality multichannel audio codecs,'' \emph{J.
  Audio Eng. Soc}, vol.~61, no. 7/8, pp. 535--545, 2013. [Online]. Available:
  \url{http://www.aes.org/e-lib/browse.cfm?elib=16869}
\BIBentrySTDinterwordspacing

\bibitem{Par2056_2019}
\BIBentryALTinterwordspacing
S.~van~de Par, S.~Disch, A.~Niedermeier, E.~Burdiel~Pérez, and B.~Edler,
  ``Temporal envelope-based psychoacoustic modelling for evaluating
  non-waveform preserving audio codecs,'' in \emph{AES Convention}, New York,
  2019, p. 10314. [Online]. Available:
  \url{http://www.aes.org/e-lib/browse.cfm?elib=20686}
\BIBentrySTDinterwordspacing

\bibitem{2fKastner}
T.~Kastner and J.~Herre, ``An efficient model for estimating subjective quality
  of separated audio source signals,'' in \emph{2019 IEEE Workshop on
  Applications of Signal Processing to Audio and Acoustics (WASPAA)}, 2019, pp.
  95--99.

\bibitem{delgado2023improved}
P.~M. Delgado and J.~Herre, ``An improved metric of informational masking for
  perceptual audio quality measurement,'' in \emph{2023 IEEE Workshop on
  Applications of Signal Processing to Audio and Acoustics (WASPAA)}, 2023, pp.
  1--5.

\bibitem{ITUP800}
{ITU-T Rec. P.800}, \emph{{Methods for subjective determination of transmission
  quality}}, Geneva, Switzerland, Aug. 1996.

\bibitem{mittag2021nisqa}
G.~Mittag, B.~Naderi, A.~Chehadi, and S.~M{\"o}ller, ``{NISQA}: A deep
  {CNN}-self-attention model for multidimensional speech quality prediction
  with crowdsourced datasets,'' \emph{arXiv preprint arXiv:2104.09494}, 2021.

\end{thebibliography}
\bibliographystyle{IEEEtran}

\vspace{11pt}

\begin{IEEEbiography}[{\includegraphics[width=1in,height=1.25in,clip,keepaspectratio]{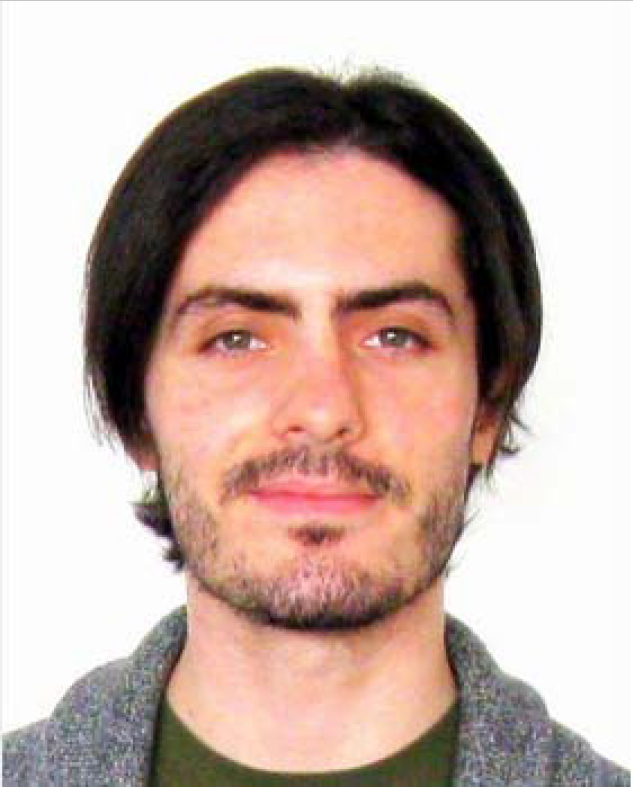}}]{Pablo M. Delgado}
    is a member of the scientific staff at the Advanced Audio Research Group, Fraunhofer Institute for Integrated Circuits (IIS) in Erlangen, Germany. He specializes in psychoacoustics applied to audio and speech coding, as well as quality assessment. He holds an Electronics Engineering degree from the University of Buenos Aires and earned his Ph.D. in psychoacoustics, machine learning, and audio signal processing from the University of Erlangen-Nuremberg under the supervision of Jürgen Herre, a key figure in the development of the MP3 and AAC audio coding formats.
   
     Dr. Delgado has contributed to the development of technologies such as the MPEG-H standard for 3D Audio Coding and extended ITU-R audio quality assessment methods like PEAQ (Perceptual Evaluation of Audio Quality). His current projects include research on six degrees of freedom (6DoF) audio for virtual reality as part of the MPEG-I standardization committee, and the use of human auditory models to inform deep learning audio tasks. He has also worked on designing, developing, and optimizing cross-platform audio coding software aligned with MPEG and 3GPP standards, as well as bit-precise implementations for low-power devices. He holds memberships in IEEE and the Audio Engineering Society (AES) and is active in the AES Technical Committees on Coding of Audio Signals, Machine Learning \& Artificial Intelligence in Audio, and Perception \& Subjective Evaluation of Audio Signals.
   \end{IEEEbiography}
\begin{IEEEbiography}[{\includegraphics[width=1in,height=1.25in,clip,keepaspectratio]{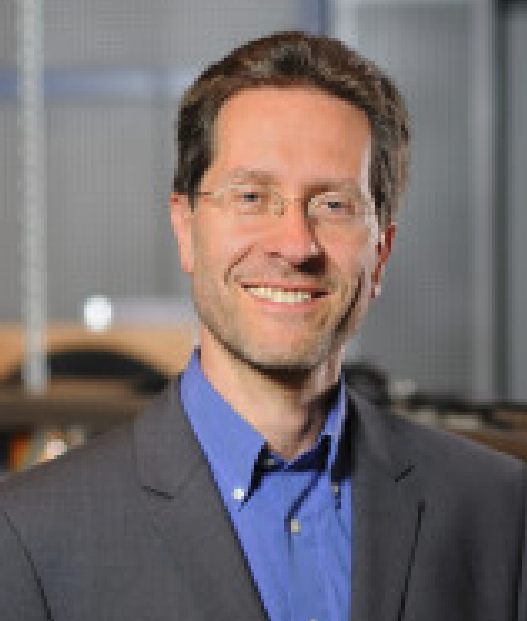}}]{Jürgen Herre}
    joined the Fraunhofer Institute for Integrated Circuits (IIS) in Erlangen, Germany, in 1989. Since then he has been involved in the development of perceptual coding algorithms for high quality audio, including the well-known ISO/MPEG-Audio Layer III coder (aka ``MP3"). In 1995, Dr. Herre joined Bell Laboratories for a Post-Doctoral term working on the development of MPEG-2 Advanced Audio Coding (AAC). By the end of '96 he went back to Fraunhofer IIS to work on the development of more advanced multimedia technology including MPEG-4, MPEG-7, MPEG-D and MPEG-I, currently as the Chief Executive Scientist for the Audio and Media Technologies division at Fraunhofer IIS, Erlangen. In September 2010, Dr. Herre was appointed professor at the University of Erlangen and the International Audio Laboratories Erlangen. 
    
    Dr. Herre is a fellow of the Audio Engineering Society, co-chair of the AES Technical Committee on Coding of Audio Signals and vice chair of the AES Technical Council. He is a senior member of the IEEE, served as a member of the IEEE Technical Committee on Audio and Acoustic Signal Processing and as an associate editor of the IEEE Transactions on Speech and Audio Processing and was a long-time member of the MPEG audio subgroup.
    
    Dr. Herre was a recipient of two Fraunhofer Awards in 1992 and 2004, respectively, and the Eduard-Rhein Award in 2015.

\end{IEEEbiography}

\vfill

\end{document}